\documentclass[twocolumn]{aastex631}
\usepackage{textcomp}
\usepackage{amsmath}
\usepackage{savesym}
\savesymbol{tablenum}
\savesymbol{figenum}
\usepackage{siunitx}
\usepackage{academicons}
\usepackage{seqsplit}
\usepackage{hyperref}
\usepackage[normalem]{ulem}
\definecolor{orcidlogocol}{HTML}{A6CE39}
\restoresymbol{SIX}{tablenum}
\usepackage{soul}

\definecolor{armygreen}{rgb}{0.29,0.33,0.13} 
\definecolor{boston}{rgb}{0.8,0.0,0.0}	
\definecolor{americanrose}{rgb}{1.0, 0.01,0.24}
\definecolor{antiquefuchsia}{rgb}{0.57, 0.36, 0.51}
\definecolor{darktangerine}{rgb}{1.0, 0.66, 0.07}

\newcommand\dagostini{D\textquotesingle Agostini}

\newcommand\daysused{2143 days}

\newcommand\daysusedDM{511 days}
\newcommand\ncpused{$ncp_{prior}=9.5$}
\newcommand\mrkbbftwone{38 blocks}
\newcommand\mrkbbfoone{5 blocks}

\newcommand\tranftwoneCorPa{$2.10  \pm 0.17$}
\newcommand\tranftwoneCorPb{$4.21 \pm 2.78$}
\newcommand\tranftwoneCorPs{$8.72  \pm 1.45$}
\newcommand\tranftwoneCorR{0.676}
\newcommand\tranftwoneCorSize{74}
\newcommand\tranftwoneCorPval{$3.93 \times 10^{-11}$}

\newcommand\tranfooneCorSize{3}

\newcommand\bbftwoneCorPa{$2.04 \pm 0.15$}
\newcommand\bbftwoneCorPb{$-6.87  \pm 1.90$}
\newcommand\bbftwoneCorPs{$5.27  \pm 0.79$}
\newcommand\bbftwoneCorR{0.892}
\newcommand\bbftwoneCorSize{25}
\newcommand\bbftwoneCorPval{$2.05\times 10^{-9}$}
\newcommand\bbfooneCorPa{$3.07 \pm 0.34$}
\newcommand\bbfooneCorPb{$-1.87  \pm 0.49$}
\newcommand\bbfooneCorPs{$0$}
\newcommand\bbfooneCorR{0.947}
\newcommand\bbfooneCorSize{5}
\newcommand\bbfooneCorPval{$1.46\times 10^{-2}$}
\newcommand\SIbbftwoneCorPa{$-3.07 \pm 0.42 \times 10^{-11}$}
\newcommand\SIbbftwoneCorPb{$99.7  \pm 10.3$}
\newcommand\SIbbftwoneCorPs{$10.7  \pm 1.4$}
\newcommand\SIbbftwoneCorR{$-0.686$}
\newcommand\SIbbftwoneCorSize{37}
\newcommand\SIbbftwoneCorPval{$2.76\times 10^{-6}$}

\newcommand\DCFMrktwone{$0.84 \pm 0.13$ }

\def\apjl{ApJ}
\def\apjs{ApJS}
\def\ao{Appl.~Opt.}
\def\aap{A\&A}

\revised{\today}
\submitjournal{ApJ}

\shorttitle{Mrk~421 and Mrk~501 Monitoring}
\graphicspath{{./}{figures/}}

\begin{document}

\title{Study of long-term spectral evolution and X-ray and Gamma-ray correlation of blazars seen by HAWC}

\author{R.~Alfaro}
\affiliation{Instituto de F\'{i}sica, Universidad Nacional Autónoma de México, Ciudad de Mexico, Mexico }

\author{C.~Alvarez}
\affiliation{Universidad Autónoma de Chiapas, Tuxtla Gutiérrez, Chiapas, México}

\author{A.~Andrés}
\affiliation{Instituto de Astronom\'{i}a, Universidad Nacional Autónoma de México, Ciudad de Mexico, Mexico }

\author{J.C.~Arteaga-Velázquez}
\affiliation{Universidad Michoacana de San Nicolás de Hidalgo, Morelia, Mexico }

\author{D.~Avila Rojas}
\affiliation{Instituto de F\'{i}sica, Universidad Nacional Autónoma de México, Ciudad de Mexico, Mexico }

\author{H.A.~Ayala Solares}
\affiliation{Department of Physics, Pennsylvania State University, University Park, PA, USA }

\author{R.~Babu}
\affiliation{Department of Physics, Michigan Technological University, Houghton, MI, USA }

\author{E.~Belmont-Moreno}
\affiliation{Instituto de F\'{i}sica, Universidad Nacional Autónoma de México, Ciudad de Mexico, Mexico }

\author{A.~Bernal}
\affiliation{Instituto de Astronom\'{i}a, Universidad Nacional Autónoma de México, Ciudad de Mexico, Mexico }

\author{K.S.~Caballero-Mora}
\affiliation{Universidad Autónoma de Chiapas, Tuxtla Gutiérrez, Chiapas, México}

\author{T.~Capistrán}
\affiliation{Instituto de Astronom\'{i}a, Universidad Nacional Autónoma de México, Ciudad de Mexico, Mexico }

\author{A.~Carramiñana}
\affiliation{Instituto Nacional de Astrof\'{i}sica, Óptica y Electrónica, Puebla, Mexico }

\author{F.~Carreón}
\affiliation{Instituto de Astronom\'{i}a, Universidad Nacional Autónoma de México, Ciudad de Mexico, Mexico }

\author{S.~Casanova}
\affiliation{Instytut Fizyki Jadrowej im Henryka Niewodniczanskiego Polskiej Akademii Nauk, IFJ-PAN, Krakow, Poland }

\author{U.~Cotti}
\affiliation{Universidad Michoacana de San Nicolás de Hidalgo, Morelia, Mexico }

\author{J.~Cotzomi}
\affiliation{Facultad de Ciencias F\'{i}sico Matemáticas, Benemérita Universidad Autónoma de Puebla, Puebla, Mexico }

\author{S.~Coutiño de León}
\affiliation{Department of Physics, University of Wisconsin-Madison, Madison, WI, USA }

\author{E.~De la Fuente}
\affiliation{Departamento de F\'{i}sica, Centro Universitario de Ciencias Exactase Ingenierias, Universidad de Guadalajara, Guadalajara, Mexico }

\author{D.~Depaoli}
\affiliation{Max-Planck Institute for Nuclear Physics, 69117 Heidelberg, Germany}

\author{N.~Di Lalla}
\affiliation{Department of Physics, Stanford University: Stanford, CA 94305–4060, USA}

\author{R.~Diaz Hernandez}
\affiliation{Instituto Nacional de Astrof\'{i}sica, Óptica y Electrónica, Puebla, Mexico }

\author{B.L.~Dingus}
\affiliation{Physics Division, Los Alamos National Laboratory, Los Alamos, NM, USA }

\author{M.A.~DuVernois}
\affiliation{Department of Physics, University of Wisconsin-Madison, Madison, WI, USA }

\author{M.~Durocher}
\affiliation{Physics Division, Los Alamos National Laboratory, Los Alamos, NM, USA }

\author{J.C.~Díaz-Vélez}
\affiliation{Department of Physics, University of Wisconsin-Madison, Madison, WI, USA }

\author{K.~Engel}
\affiliation{Department of Physics, University of Maryland, College Park, MD, USA }

\author{C.~Espinoza}
\affiliation{Instituto de F\'{i}sica, Universidad Nacional Autónoma de México, Ciudad de Mexico, Mexico }

\author{K.L.~Fan}
\affiliation{Department of Physics, University of Maryland, College Park, MD, USA }

\author{N.~Fraija}
\affiliation{Instituto de Astronom\'{i}a, Universidad Nacional Autónoma de México, Ciudad de Mexico, Mexico }

\author{S.~Fraija}
\affiliation{Instituto de Astronom\'{i}a, Universidad Nacional Autónoma de México, Ciudad de Mexico, Mexico }

\author{J.A.~Garc\'{i}a-Gonz\'{a}lez\correspondingauthor}
\affiliation{Tecnologico de Monterrey, Escuela de Ingenier\'ia y Ciencias, Ave. Eugenio Garza Sada 2501, Monterrey, N.L., Mexico, 64849}
\email{anteus79@tec.mx}

\author{F.~Garfias}
\affiliation{Instituto de Astronom\'{i}a, Universidad Nacional Autónoma de México, Ciudad de Mexico, Mexico }

\author{A.~Gonzalez Muñoz}
\affiliation{Instituto de F\'{i}sica, Universidad Nacional Autónoma de México, Ciudad de Mexico, Mexico }

\author{M.M.~González\correspondingauthor}
\affiliation{Instituto de Astronom\'{i}a, Universidad Nacional Autónoma de México, Ciudad de Mexico, Mexico }
\email{magda@astro.unam.mx}
\author{J.A.~Goodman}
\affiliation{Department of Physics, University of Maryland, College Park, MD, USA }

\author{S.~Groetsch}
\affiliation{Department of Physics, Michigan Technological University, Houghton, MI, USA }

\author{J.P.~Harding}
\affiliation{Physics Division, Los Alamos National Laboratory, Los Alamos, NM, USA }

\author{S.~Hernández-Cadena}
\affiliation{Instituto de Física, Universidad Nacional Autónoma de México, Ciudad de Mexico, Mexico }

\author{I.~Herzog}
\affiliation{Department of Physics and Astronomy, Michigan State University, East Lansing, MI, USA }

\author{D.~Huang}
\affiliation{Department of Physics, University of Maryland, College Park, MD, USA }

\author{F.~Hueyotl-Zahuantitla}
\affiliation{Universidad Autónoma de Chiapas, Tuxtla Gutiérrez, Chiapas, México}

\author{A.~Iriarte}
\affiliation{Instituto de Astronom\'{i}a, Universidad Nacional Autónoma de México, Ciudad de Mexico, Mexico }

\author{V.~Joshi}
\affiliation{ECAP}

\author{S.~Kaufmann}
\affiliation{Universidad Politecnica de Pachuca, Pachuca, Hgo, Mexico }

\author{D.~Kieda}
\affiliation{Department of Physics and Astronomy, University of Utah, Salt Lake City, UT, USA }

\author{A.~Lara}
\affiliation{Instituto de Geof\'{i}sica, Universidad Nacional Autónoma de México, Ciudad de Mexico, Mexico }

\author{W.H.~Lee}
\affiliation{Instituto de Astronom\'{i}a, Universidad Nacional Autónoma de México, Ciudad de Mexico, Mexico }

\author{J.~Lee}
\affiliation{UOS}

\author{H.~León Vargas}
\affiliation{Instituto de F\'{i}sica, Universidad Nacional Autónoma de México, Ciudad de Mexico, Mexico }

\author{J.T.~Linnemann}
\affiliation{Department of Physics and Astronomy, Michigan State University, East Lansing, MI, USA }

\author{A.L.~Longinotti}
\affiliation{Instituto de Astronom\'{i}a, Universidad Nacional Autónoma de México, Ciudad de Mexico, Mexico }

\author{G.~Luis-Raya}
\affiliation{Universidad Politecnica de Pachuca, Pachuca, Hgo, Mexico }

\author{K.~Malone}
\affiliation{Physics Division, Los Alamos National Laboratory, Los Alamos, NM, USA }

\author{O.~Martinez}
\affiliation{Facultad de Ciencias F\'{i}sico Matemáticas, Benemérita Universidad Autónoma de Puebla, Puebla, Mexico }

\author{I.~Martinez-Castellanos}
\affiliation{NASA Goddard Space Flight Center, Greenbelt, MD 20771, USA  }

\author{J.~Martínez-Castro}
\affiliation{Centro de Investigaci\'{o}n en Computaci\'{o}n, Instituto Polit\'{e}cnico Nacional, M\'{e}xico City, M\'{e}xico.}

\author{J.A.~Matthews}
\affiliation{Dept of Physics and Astronomy, University of New Mexico, Albuquerque, NM, USA }

\author{P.~Miranda-Romagnoli}
\affiliation{Universidad Autónoma del Estado de Hidalgo, Pachuca, Mexico }

\author{J.A.~Montes}
\affiliation{Instituto de Astronom\'{i}a, Universidad Nacional Autónoma de México, Ciudad de Mexico, Mexico }

\author{E.~Moreno}
\affiliation{Facultad de Ciencias F\'{i}sico Matemáticas, Benemérita Universidad Autónoma de Puebla, Puebla, Mexico }

\author{M.~Mostafá}
\affiliation{Department of Physics, Pennsylvania State University, University Park, PA, USA }

\author{A.~Nayerhoda}
\affiliation{Instytut Fizyki Jadrowej im Henryka Niewodniczanskiego Polskiej Akademii Nauk, IFJ-PAN, Krakow, Poland }

\author{L.~Nellen}
\affiliation{Instituto de Ciencias Nucleares, Universidad Nacional Autónoma de Mexico, Ciudad de Mexico, Mexico }

\author{M.U.~Nisa}
\affiliation{Department of Physics and Astronomy, Michigan State University, East Lansing, MI, USA }

\author{R.~Noriega-Papaqui}
\affiliation{Universidad Autónoma del Estado de Hidalgo, Pachuca, Mexico }

\author{N.~Omodei}
\affiliation{Department of Physics, Stanford University: Stanford, CA 94305–4060, USA}

\author{M.~Osorio}
\affiliation{Instituto de Astronom\'{i}a, Universidad Nacional Autónoma de México, Ciudad de Mexico, Mexico }

\author{Y.~Pérez Araujo}
\affiliation{Instituto de F\'{i}sica, Universidad Nacional Autónoma de México, Ciudad de Mexico, Mexico }
\affiliation{Instituto de Astronom\'{i}a, Universidad Nacional Autónoma de México, Ciudad de Mexico, Mexico }

\author{E.G.~Pérez-Pérez}
\affiliation{Universidad Politecnica de Pachuca, Pachuca, Hgo, Mexico }

\author{C.D.~Rho}
\affiliation{SKKU}

\author{D.~Rosa-González}
\affiliation{Instituto Nacional de Astrof\'{i}sica, Óptica y Electrónica, Puebla, Mexico }

\author{E.~Ruiz-Velasco}
\affiliation{Max-Planck Institute for Nuclear Physics, 69117 Heidelberg, Germany}

\author{H.~Salazar}
\affiliation{Facultad de Ciencias F\'{i}sico Matemáticas, Benemérita Universidad Autónoma de Puebla, Puebla, Mexico }

\author{D.~Salazar-Gallegos}
\affiliation{Department of Physics and Astronomy, Michigan State University, East Lansing, MI, USA }

\author{A.~Sandoval}
\affiliation{Instituto de F\'{i}sica, Universidad Nacional Autónoma de México, Ciudad de Mexico, Mexico }

\author{M.~Schneider}
\affiliation{Department of Physics, University of Maryland, College Park, MD, USA }

\author{J.~Serna-Franco}
\affiliation{Instituto de F\'{i}sica, Universidad Nacional Autónoma de México, Ciudad de Mexico, Mexico }

\author{A.J.~Smith}
\affiliation{Department of Physics, University of Maryland, College Park, MD, USA }

\author{Y.~Son}
\affiliation{UOS}

\author{R.W.~Springer}
\affiliation{Department of Physics and Astronomy, University of Utah, Salt Lake City, UT, USA }

\author{O.~Tibolla}
\affiliation{Universidad Politecnica de Pachuca, Pachuca, Hgo, Mexico }

\author{K.~Tollefson}
\affiliation{Department of Physics and Astronomy, Michigan State University, East Lansing, MI, USA }

\author{I.~Torres}
\affiliation{Instituto Nacional de Astrof\'{i}sica, Óptica y Electrónica, Puebla, Mexico }

\author{R.~Torres-Escobedo}
\affiliation{SJTU}

\author{R.~Turner}
\affiliation{Department of Physics, Michigan Technological University, Houghton, MI, USA }

\author{F.~Ureña-Mena}
\affiliation{Instituto Nacional de Astrof\'{i}sica, Óptica y Electrónica, Puebla, Mexico }

\author{E.~Varela}
\affiliation{Facultad de Ciencias F\'{i}sico Matemáticas, Benemérita Universidad Autónoma de Puebla, Puebla, Mexico }

\author{L.~Villaseñor}
\affiliation{Facultad de Ciencias F\'{i}sico Matemáticas, Benemérita Universidad Autónoma de Puebla, Puebla, Mexico }

\author{X.~Wang}
\affiliation{Department of Physics, Michigan Technological University, Houghton, MI, USA }

\author{I.J.~Watson}
\affiliation{UOS}

\author{K.~Whitaker}
\affiliation{Department of Physics, Pennsylvania State University, University Park, PA, USA }

\author{E.~Willox}
\affiliation{Department of Physics, University of Maryland, College Park, MD, USA }

\author{S.~Yun-Cárcamo}
\affiliation{Department of Physics, University of Maryland, College Park, MD, USA }

\author{H.~Zhou}
\affiliation{SJTU}

\author{C.~de León}
\affiliation{Universidad Michoacana de San Nicolás de Hidalgo, Morelia, Mexico }

\collaboration{100}{(HAWC collaboration)}

\author{Abraham D. Falcone}
\affiliation{Pennsylvania State University, Astronomy \& Astrophysics Dept., University Park, PA}

\author{Fredric Hancock}
\affiliation{Pennsylvania State University, Astronomy \& Astrophysics Dept., University Park, PA}
\affiliation{now at University of Illinois Urbana Champaign, Dept. of Physics, Urbana, IL}
\begin{abstract}

The HAWC Observatory collected 6 years of extensive data, providing an ideal platform for long-term monitoring of blazars in the Very High Energy (VHE) band, without bias towards specific flux states. HAWC continuously monitors blazar activity at TeV energies, focusing on sources with a redshift of $z\leq0.3$, based on the Third \textit{Fermi}-LAT Catalog of High-Energy sources. We specifically focused our analysis on Mrk~421 and Mrk~501, as they are the brightest blazars observed by the HAWC Observatory. With a dataset of \daysused~, this work significantly extends the  monitoring previously published, which was based on \daysusedDM~of observation. By utilizing HAWC data for the VHE $\gamma$-ray emission in the 300 GeV–100 TeV energy range, in conjunction with Swift-XRT data for the 0.3 to 10 keV X-ray emission, we aim to explore potential correlations between these two bands. For Mrk~501, we found evidence of a long-term correlation. Additionally, we identified a period in the light curve where the flux was very low for more than two years. On the other hand, our analysis of Mrk~421 measured a strong linear correlation for quasi-simultaneous observations collected by HAWC and Swift-XRT. This result is consistent with a linear dependence and a multiple-zone synchrotron self-Compton model to explain the X-ray and the $\gamma$-ray emission. Finally, as suggested by previous findings, we confirm a harder-when-brighter behavior in the spectral evolution of the flux properties for Mrk~421. These findings contribute to the understanding of blazar emissions and their underlying mechanisms.

\end{abstract}



\section{Introduction} \label{sec:intro}
Blazars are active galactic nuclei (AGN) characterized by the presence of a relativistic jet closely aligned with the observer's line of sight, typically with estimated viewing angles of $\theta\lesssim 10^{\circ}$. These objects exhibit high variability across the entire electromagnetic spectrum. Extensive research has explored the variability in blazars,  revealing both short and long-timescale variations \citep{1997ARA&A..35..445U,refId01,2014A&ARv..22...73F,Acciari_2020b}.

Blazars have a spectral energy distribution (SED) that exhibits a distinctive double peak structure. Based on the location of the synchrotron peak of the low-energy component of their SED, they can be classified into three groups: low-energy peaked BL Lac (LBL), intermediate-energy peaked BL Lac (IBL) and high-energy peaked BL Lac (HBL) \citep{1995ApJ...444..567P}. It is well established that the low energy peak comes from synchrotron radiation generated by relativistic electrons within the jet~\citep{ulrich_1997}

In the SED of Blazars, a second peak in the high-energy component is observed at $\gamma$-ray energies and sometimes includes measurable TeV emission, primarily attributed to Synchrotron Self-Compton (SSC) mechanism. The influence of hadronic contributions on this emission remains relatively unknown.

Correlations between TeV $\gamma$-rays and X-rays have been observed on multiple occasions, both during high activity states or flares \citep{1995ApJ...449L..99M,2004NewAR..48..419F, 2007ApJ...663..125A,2019MNRAS.484.2944G, Acciari_2021} and also during quiescent states \citep{2015A&A...576A.126A, 2016ApJ...819..156B,Acciari_2020b}. In only a few cases, TeV $\gamma$-rays flares without a corresponding X-ray flaring emission have been detected,  referred to as ``orphan flares'' \citep{2005ApJ...630..130B,2009ApJ...703..169A}.

The interpretation of correlations found between optical bands and TeV $\gamma$-rays/X-rays remains controversial. Some studies have reported optical/TeV and $\gamma$-ray/X-ray correlations, as well as non-correlations, with different time lags~\citep{2009ApJ...695..596H, 2015A&A...576A.126A} while others have found no correlations at all~\citep{1995ApJ...449L..99M,2015A&A...576A.126A,2007ApJ...663..125A}.
Similarly, the emission in the radio band does not typically appear to be  correlated with TeV $\gamma$-rays~\citep{2011ApJ...738...25A}.

Detailed studies of the correlation between X-rays and TeV emissions are crucial for constraining and distinguishing between leptonic and hadronic models of the SED. Mrk~501 has been extensively investigated in both quiescent and flaring states with the aim of exploring correlations between the TeV $\gamma$-ray, X-ray, and optical bands on timescales ranging from minutes to days \citep{2006ApJ...646...61G, 2017MNRAS.469.1655K, 2019ApJ...870...93A, 2019NewA...7301278S, 2020Galax...8...55P, 2017MNRAS.469.1655K,2006ApJ...646...61G}.

For example,  a direct correlation between X-ray and TeV $\gamma$-ray emissions was confirmed during the years 1997, 1998, 1999, 2000, and 2004, which appeared to be stronger when the source was brighter~\citep{2006ApJ...646...61G}. In 2014, from March to October, there was no significant correlation between emissions in the energy band of 0.3–300 GeV and optical-UV fluxes. However, a strong correlation was found between the TeV $\gamma$-ray and X-ray during this period~\citep{2017MNRAS.469.1655K}. In a more recent study, a trend of harder-when-brighter behavior was observed for Mrk~421~\citep{Acciari_2021}. However, despite the campaign spanning several years, it primarily featured low-activity periods and did not lead to a definitive conclusion. 

HAWC conducts surveys of the sky to monitor blazar activity at TeV energies~\citep{Albert_2021}. For this search, HAWC utilizes the Third Fermi-LAT Catalog of High-Energy sources~\citep{Ajello_2017}, targeting sources with a redshift of $z\leq0.3$. In this context, HAWC has successfully detected two of the brightest blazars, Mrk~421 and Mrk~501.

These sources were previously studied in detail in a prior work~\citep{Abeysekara_2017a} where the light curves for a total of \daysusedDM~ were presented.  In an attempt to identify a possible X-ray/$\gamma$-ray correlation, \textit{Swift}-BAT data in the 15-50 keV band for the X-ray emission and HAWC data for the very high-energy (VHE) emission were examined. However, no significant evidence of a correlation between the two bands was found. Additionally, the study reported that the average flux for  Mrk~421 in the same period of time was approximately 0.8 Crab Units (CU) above 1 TeV. Mrk~501 exhibited an average flux of approximately 0.3~CU above 1 TeV.

In this paper, we extend our analysis utilizing data from Nov 2014 to Oct 2020 for a total duration of \daysused. We employ \textit{Swift}-XRT data in the 0.3-10 keV band for the X-ray emission.

Furthermore, if the $\gamma$-ray emission at TeV energies arises from the SSC mechanism, the corresponding synchrotron emission would occur at softer X-rays compared to what is observed by \textit{Swift}-BAT. In consequence, this synchrotron emission would be closer to its peak, ranging from 0.3 to 5 keV~\citep{Bartoli_2016}, which corresponds to the energy range of \textit{Swift}-XRT.

The primary objective of this work is to use data from long-term monitoring to investigate whether a correlation exists between TeV emission and soft X-ray emission. Such a correlation 
is a natural consequence of the SSC model as the underlying $\gamma$-ray emission mechanism for Mrk 421 and Mrk~501. If confirmed, we can also assess the harder-when-brighter evolution of the spectrum. A detailed description of this correlation and their robustness may suggest contributions from other processes, or may provide insights into the emission's mechanisms, such as the number of emission zones~\citep{Katarzyski2010} or whether the emission occurs in the Klein-Nishina limit~\citep{Tavecchio_1998}.

We present our results as follows: in Section~\ref{sec:sources} we provide a brief overview of the two sources of interest, including details about the observation campaigns and the results of X-ray/$\gamma$-ray correlation analysis. Section~\ref{sec:analysisHAWC} presents an overview of the HAWC Observatory and outlines our data analysis process. Section~\ref{sec:analysisXRT} briefly covers the reduction of \textit{Swift}-XRT data. In Section~\ref{sec:bb} we extend our Bayesian blocks analysis to the data sample of \daysused~for both Mrk~501 and Mrk~421. Section~\ref{sec:xraycorr} introduces the X-ray/$\gamma$-ray correlation and outlines our methodology for assessing result robustness. In Section~\ref{sec:gammaVsindex} we measure the $\gamma$-ray vs. photon index correlation for Mrk~421. Finally, Section~\ref{sec:conclusions} discusses the results and conclusions obtained in this study. 

\section{The sources}\label{sec:sources}
HAWC observes two of the brightest blazars in the northern TeV sky, which have been the focus of multiple observation campaigns. Extensive monitoring of Mrk~421 and Mrk~501 has significantly enhanced our understanding of their broadband emission, as summarized in the following two subsections.
\subsection{Mrk~421}\label{sec:SourceMrk421}

VHE emission from Mrk~421 (z=$0.031$) was first detected by the Whipple Observatory~\citep{Punch:1992xw}. Extensive monitoring campaigns has been conducted at various activity levels. Some campaigns were not restricted to specific source states and included both low and high states in the data analysis. For instance,  one campaign utilized data from the Whipple 10-meter telescope for the TeV $\gamma$-ray emission, alongside data from Rossi X-Ray Timing Explorer (RXTE). This campaign, conducted from 2003 to 2004~\citep{Blazejowski_2005}, yielded valuable insights, including variability studies and correlation between the two emissions. Notably, it revealed the occurrence of TeV flares with no counterparts at longer wavelengths. According to the authors, the analysis of these flares during that period posed challenges to both leptonic and hadronic models.


Another example of an intermediate flux state campaign occurred over two months in December 2002 and January of 2003~\citep{Rebillot_2006}.

Regarding low-activity states, a campaign spanning 4.5 months in 2009 ~\citep{refId0} deserves mention. During this period,  when the average flux of Mrk~421 was approximately 0.5 times the flux of the Crab Nebula, the campaign successfully identified a strong X-ray/$\gamma$-ray correlation. This correlation was established using data from RXTE/PCA and \textit{Swift}-XRT 2–10 keV) for X-rays, and MAGIC, Whipple for VHE $\gamma$-rays. Additionally, the campaign observed a trend of harder-when-brighter behavior in the X-ray spectra.

Multiple campaigns have been conducted to investigate high-activity states of the source. One illustrative example is a campaign during the 2006 and 2008 outburst of Mrk~421~\citep{Acciari_2009}. This multi-wavelength campaign incorporated UV and X-ray data from \textit{XMM-Newton} along with $\gamma$-ray data from the 10 m Whipple telescope, VERITAS and MAGIC telescopes. A primary focus was to obtain simultaneous data across these three wavelengths. The 2006 data facilitated the study of source variability, while the 2008 data aimed to identify correlations between X-ray and $\gamma$-ray bands. However,  the study reported not measurable correlation in the sample used.

Regarding the measurement of X-ray/$\gamma$-ray correlation at TeV energies, several analyses have identified correlation between the two bands~\citep{Fossati_2008,Bartoli_2016,ACCIARI20141,Acciari_2020b,2021A&A...655A..89M} utilizing either GeV or TeV band for the $\gamma$-ray emission.

In some of the aforementioned studies, evidence supports a harder-when-brighter trend for the spectra. However, it is important to note that in certain cases, this behavior can not be extended to the highest fluxes~\citep{Acciari_2020b}.

The HAWC Collaboration recently conducted a time-averaged analysis of the TeV SED for Mrk~421 using a data set spanning 1038 days. Their analysis revealed that the intrinsic spectrum of this source can be adequately represented by a power law with an exponential cut-off~\citep{Albert_2022a}. Additionally, this analysis determined that the maximum energy at which the source is detected is ~ 9 TeV.

In this work, we build upon previous monitoring efforts of Mrk~421 which utilized \daysusedDM~of data and initially explored the possibility of X-ray/$\gamma$-ray correlation using \textit{Swift}-BAT data. However,  no conclusive evidence of such correlation was found in that preliminary investigation~\citep{Abeysekara_2017a}. One potential explanation for the absence of correlation, as discussed in Section~\ref{sec:intro}, may relate to the expectation in the SSC model that the $\gamma$-ray emission  is strongly linked to a significantly softer X-ray component than that traced by \textit{Swift}-BAT, which has a hard band pass at $E>15$ keV, providing therefore a strong motivation for  the use of \textit{Swift}-XRT data in these studies~\citep{Bartoli_2016}.
\subsection{Mrk~501}\label{sec:SourceMrk501}
VHE emission from Mrk~501 ($z=0.034$) was first detected by the Whipple Observatory~\citep{Quinn_1996}.

In the study performed by~\citet{refId0_2010}, a comprehensive analysis covering over a decade of data collected from various experiments revealed a X-ray/$\gamma$-ray linear correlation for Mrk~501. It is important to note that the flux measurements were not strictly simultaneous, as they were averaged over 12 hrs.

During an observation campaign conducted from October to November of 2011~\citep{Bartoli_2012}, the ARGO-YBJ experiment reported no significant correlation between X-ray emission in the 15-50 keV band using \textit{Swift}-BAT data  and $\gamma$-ray emission ($> 0.3$ TeV). The source was also analyzed during the X-ray flaring periods and although there was an increase in the $\gamma$-ray emission, there was no evidence of correlation with X-ray emission.
 
In a high activity campaign~\citep{refId0_mrk501}, a significant correlation between the two bands was reported, and this correlation appeared to increase slightly with higher X-ray energies. It is noteworthy that during the low activity periods, the correlation weakened, and in some cases, it disappeared.

For long-term measurements, a campaign spanning over 5 years found that the correlation remain strong and could exhibit both linear and nonlinear behaviours~\citep{Gliozzi_2006}. Additionally, the study characterized the long-term  X-ray flux and spectral variability. In terms of long-term timescales, the spectrum tended to harden when the source brightened, consistent with typical blazar behavior.

Multi-wavelength correlations spanning from 2012 to 2018 were also studied by\citet{2021A&A...655A..93A} and a strong TeV/X-ray correlation with zero lag was  found.

HAWC also performed a time averaged analysis of the TeV SED for Mrk~501~\citep{Albert_2022a}, and it found that its intrinsic spectra can be represented with a simple power law. It also measured that the maximum energy at which the source is detected is 12 TeV.

It is noteworthy to mentioned that Mrk~501 has shown a low broadband activity for an extended period of time, starting at mid 2017 an extending up to 2019, as reported by \citet{2023ApJS..266...37A}. where a significant correlation between X-ray and $\gamma$-ray was also found.

\section{Analysis}\label{sec:analysis}
\subsection{HAWC data analysis}\label{sec:analysisHAWC}
HAWC is comprised of an array of 300 water-Cherenkov detectors (WCDs), each with dimensions of 7.3 meters in diameter and 4.5 meters in depth~\citep{ABEYSEKARA2023168253}. The total effective area is approximately $\rm 22,000\ m^2$ for the 300 WCDs. Situated at an elevation of approximately 4,100 meters above sea level, HAWC is located on the slopes of the Sierra Negra volcano in the state of Puebla, Mexico. It has been in operation for more than seven years, boasting an impressive duty cycle exceeding 95\%.

With an instantaneous field of view of $\rm 2\ sr$, HAWC surveys two-thirds of the sky every day, operating with an energy range spanning from $\rm 300\ GeV$ to $\rm 100\ TeV$. One of the key advantages of the HAWC Observatory is its ability to provide continuous and unbiased monitoring, unrestricted by seasonal observing limitations. Since late 2014, it has monitored the TeV sky and accumulated a vast data sample spanning over 6 years.



For this analysis, we utilized the same data reconstruction version as employed in ~\citet{Abeysekara_2017a}. The dataset covers the period from Nov 2014 to Oct 2020. Here, we provide only pertinent details of the adopted analysis; for a comprehensive description,  please refer to \citet{Abeysekara_2017a} and the work by~\citet{Abeysekara_2017b}.

In the HAWC Observatory's event recording, a HAWC event is registered when a minimum number of photo-multiplier tubes (PMTs) trigger within a specified time window. These events are classified into nine distinct analysis bins (designated as $\mathcal{B}$ identifiers ranging from 1 to 9). The classification is based on the fraction of available PMTs triggered by each event (see ~\citet{Crab100tev} for detailed information). The Gamma/Hadron separation criteria,  outlined in \citet{Abeysekara_2017b}, are applied from bin 1 to bin 9.

To obtain the flux normalization, we employ the daily maps in sidereal days, as explained in \citet{Abeysekara_2017a}, which encompass  a full transit of a given source within the Right Ascension (RA) range of $45^{\circ} < RA < 315^{\circ}$. These daily maps were utilized for daily monitoring of blazars such as Mrk~421, Mrk~501 and the Crab Nebula~\citep{Abeysekara_2017a} and covered \daysusedDM~of data. Notably, the sources of interest transit the sky above HAWC for approximately 6.2 hr each sidereal day. To ensure accuracy in the flux normalization, we impose a minimum coverage requirement of $50\%$ of the transit, along with a condition that the zenith angle $\theta$ is less than  $45^{\circ}$. Additionally, a quality cut entails the removal of sub-runs with a rate deviation exceeding $10\sigma$ from the sinusoidal fit to the rate within a 12-hour time window.

In cases where the coverage of a source transit is interrupted due to various circumstances, such as detector downtime, the quantification of signal loss in comparison to a full transit is determined by excluding the gap period from the integration over zenith angles~\citep{Abeysekara_2017a}. Subsequently, the expected event count is adjusted accordingly. One limitation of this approach is its dependency on the declination, which requires the assumption of a minimum exposure equivalent to one full transit of the source. Consequently, this method may pose challenges when attempting to obtain results involving time scales of hours, such as intra-day variability studies or the observation of spectral hardening  during periods of high activity, which are characteristic of $\gamma$-ray flares in blazars, as in our case of study.

One significant difference between the present work and the methods described in \citet{Abeysekara_2017a} is the software employed for calculating the flux normalization. Instead of using the Likelihood Fitting Framework (LiFF)~\citep{2015ICRC...34..948Y} to obtain the
flux normalization, we adopt the Zenith Band Response Analysis (ZEBRA)~\citep{2019arXiv190806122M}, which is the latest software used for generating light curves (LCs) in HAWC analyses and offers several improvements.

LiFF relies on assumed spectra for a given source, convolving them with a detector response function that includes the point spread function (PSF). It subsequently employs the log-likelihood function to calculate the likelihood ratio test (TS) between a null hypothesis, assuming background-only, and an alternative hypothesis that uses the input flux for the source.

In contrast, ZEBRA~\citep{2019arXiv190806122M} determines the detector response's dependence on the zenith angle by means of simulation.This response is then convolved with the source's spectrum to estimate the observed counts over a specified time period. Importantly, ZEBRA eliminates the constraint of requiring a minimum exposure equivalent to one full transit.

A consistency check between both software packages was conducted for the Crab Nebula and the blazars Mrk~421 and Mrk~501 in~\citet{2019arXiv190809452G}.

We employ direct integration \citep{Atkins_2003} to derive a background estimate. In this procedure, a local efficiency map is constructed by averaging counts within strip of pixels spanning 24 hours in right ascension around any given location. Subsequently, this efficiency map is smoothed using a spline fit to mitigate the limitations posed by limited statistics in higher analysis bins.

To prevent any bias in the results, pixels near the strongest known sources and the galactic plane are excluded during the averaging process. This exclusion is necessary to avoid counting $\gamma$-ray events as part of the background. The estimated background counts for each pixel are then stored in a second map.

We consider for both sources, Mrk~421 and Mrk~501,  the spectral model as a power law with an exponential cut off given as,

\begin{equation}\label{equ:SED}
    \frac{dN}{dE} = N_{0}\left( \frac{E}{E_{0}}\right)^{-\alpha}\exp\left(\frac{E}{E_{c}}\right)
\end{equation}{}

Here, $N_{0}(\rm {TeV}^{-1}cm^{-2}s^{-1})$ represented the flux normalization, $\alpha$ is the photon index, and $E_c$(TeV) is the energy cut-off. A detailed study conducted by HAWC that takes into account the EBL absorption can be found in~\citet{Albert_2022a}

In Equation~\ref{equ:SED} the parameters $E_{0}$(TeV) (pivot energy) and $\alpha$ were fixed, with only the normalization $N_0$ as a free parameter. The specific values used for each source are summarised in Table~\ref{tab:SEDpar} and can be found in detail in~\citet{Abeysekara_2017a}.
\begin{table}[ht]
    \centering
    \begin{tabular}{c|c|c|c}
        Source & $E_{0}$(TeV) & $\alpha$ & $E_{c}$(TeV) \\
        \hline
        Mrk~421 & 1 & 2.2 & 5\\
        Mrk~501 & 1 & 1.6 & 6
    \end{tabular}
    \caption{Parameters used in the cut-off power law for the observed spectra of Mrk~421 and Mrk~501}
    \label{tab:SEDpar}
\end{table}


 The systematic uncertainties related to the calculation of the flux normalization have been extensively discussed by~\citet{Abeysekara_2017b}. The resulting uncertainty on individual flux values under spectral hardening or softening, was obtained using simulated $\gamma$-ray fluxes with different spectral parameter values. The range used for the spectral index was 2.0 to 2.4 and 1.2 to 2.0 for Mrk~421 and Mrk~501 respectively, resulting in a value of $\pm 5\%$.
 In the case of Mrk~421, the spectral index range can vary outside of the range presented here, as described in~\citet{2021A&A...655A..89M} (from 1.95 to 2.80). For Mrk~501, the range can also include values from 2.35 to 2.67~\citep{2023ApJS..266...37A}.
 Other notable sources of uncertainty are the charge resolution, angular resolution, and late light simulation, the last one being the most significant contributor to uncertainty, with a maximum impact of up to 40\%.

\subsection{\textit{Swift}-XRT data analysis}\label{sec:analysisXRT}
\textit{Swift}-XRT~\citep{swift_xrt} is an X-ray Telescope on board the Neil Gehrels Swift Observatory~\citep{Gehrels_2004}. \textit{Swift}-XRT observations of both Mrk~421 and Mrk~501, were conducted in two different modes:  the Window Timing (WT) mode and the photon counting (PC) mode. The majority of the observations were carried out in the WT mode due to the high count rates, which can lead to photon pile-up in the PC mode.

For this study, the \textit{Swift} data analyzed for both Mrk~421 and Mrk~501 spans from November 2014 to October 2020. The \textit{Swift} data were binned to match the time period of a sidereal day, corresponding to HAWC data.

Data products were extracted using tools from HEASoft\footnote{\href{https://heasarc.gsfc.nasa.gov/docs/software/lheasoft/}{\seqsplit{https://heasarc.gsfc.nasa.gov/docs/software/lheasoft/}}}. Initially, the  cleaned level-3 event files were separated into individual snapshots, each representing a pointed observation. An image was extracted for each snapshot within the energy range of  0.3-10 keV.

In the WT mode, the first 150 seconds of data were discarded from each observation to exclude data affected by spacecraft settling issues. Pile-up correction was applied using the method described in~\citet{refId01}.  For the spectrum extraction, a $40 \times 20$ pixel (2.36 arcsec/pixel) box was chosen as the source extraction region,  and an annular boxed region with dimensions $100 \times 20$ pixels was used for background spectrum extraction.

In the Photon Counting (PC) mode, a circular source region with a typical size of 20 pixels and a surrounding annular background
regionwith an inner and outer radii of typically 40 and 80 pixels,
respectively, were selected for spectra extraction.  The size was
dynamically chosen, as a function of source count rate in the region. If
the count rate is above the pile-up threshold, then an annular source
region is used. This pile-up correction was applied when the source
counts exceeded 0.6 counts/s. For further description of these methods,
refer to \citet{Stroh_2013}. Specifically, an
annular region in the center of the source region was excluded to
eliminate pile-up affected pixels, with this annular region increasing
in size for higher count rates until the pile-up becomes negligible,
below the 0.6 count/s threshold, outside of the central annulus ring.
Subsequently, a point spread function correction was applied based on
these regions, including the correction for the chosen annulus.

All spectra were binned using 1 sidereal-day intervals. Model fitting was performed using XSPEC\footnote{\href{https://heasarc.gsfc.nasa.gov/xanadu/xspec/}{\seqsplit{https://heasarc.gsfc.nasa.gov/xanadu/xspec/}}} ver 11 (Arnaud 1996) with a model consisting of a log parabola (logpar) combined with Galactic absortion as specified in the Tuebingen-Boulder ISM absorption model (tbabs). The final model took the form: \emph{cflux*tbabs*logpar}.
The \emph{cflux} component provides the unabsorbed flux from each source in the energy band of $0.3-10$ keV. The hydrogen column density along the line of sight of the two blazars was fixed at $\rm 1.9\times^{20}cm^{-2}$ for Mrk~421 and $\rm 1.7\times^{20}cm^{-2}$ for Mrk~501. These values were obtained using the HEASoft tool called Hydrogen density column calculator~\citep{refId0}\footnote{\href{https://heasarc.gsfc.nasa.gov/cgi-bin/Tools/w3nh/w3nh.pl}{https://heasarc.gsfc.nasa.gov/cgi-bin/Tools/w3nh/w3nh.pl}}
\subsection{\texorpdfstring{X-ray/$\gamma$-ray  }~Correlation}\label{sec:xraycorr}

After obtaining the $\gamma$-ray and X-ray fluxes as described in Sections~\ref{sec:analysisHAWC} and~\ref{sec:analysisXRT}, we employ the maximum likelihood method, as discussed in~\citet{dagostini1,2003RPPh...66.1383D}, to assess the linear correlation between them. This method allows us to not only determine the strength of the correlation but also measure the degree of dispersion in the data using the parameter $\sigma_{d}$.

The underlying assumption of this method is that when the correlated data pairs $(F_{x}^{i}, F_{\gamma}^{i} )$ follows a linear dependence $F_{\gamma} = aF_{x} + b$ with an intrinsic scatter $\sigma_{d}$, we can derive the optimal values of the parameters (a, b, and $\sigma_{d}$) using the likelihood function. This function takes into account the uncertainties with $F_{x}^{i}$ and $F_{\gamma}^{i}$ ($\sigma_{F_{x}^{i}}$ and $\sigma_{F_{\gamma}^{i}}$, respectively) within its calculations and can be expressed as follows:

\begin{multline}
    L(a, b, \sigma_{d}) = 
    \frac{1}{2}\sum_{i} log( \sigma_{d}^{2} + \sigma_{F_{\gamma}^{i}}^{2}+ a\sigma_{F_{x}^{i}}^{2})\\
    + \frac{1}{2}\sum_{i}\frac{(F_{\gamma^{i}}^{2} - aF_{{x}^{i}}^{2} - b^{2})^{2}}{ \sigma_{d}^{2} + \sigma_{F_{\gamma}^{i}}^{2}+ a\sigma_{F_{x}^{i}}^{2}}
\end{multline}\label{eq:dagostinie}

This approach allows us to rigorously assess the correlation between the $\gamma$-ray and X-ray fluxes, considering both their values and associated uncertainties.
\subsection{Bayesian Blocks\label{sec:bb}}
To assess the origin of the intrinsic dispersion $\sigma_{d}$ we also evaluate the flux correlation by segmenting the data into Bayesian Blocks (BB) that represent periods of time where the flux is consistent with being constant within the statistical uncertainties. In this case, we apply the BB algorithm to the HAWC fluxes and combine the X-ray fluxes accordingly.

We utilized the ``point measurements fitness function" for the Bayesian blocks algorithm, as outlined in Section~3.3 of~\citet{bblocks}, to analyze the daily flux data in the $\gamma$-ray LCs.

To apply this algorithm effectively, an initial Bayesian prior, denoted as \textit{ncp$_{prior}$}, must by chosen.  The probability of encountering a new transition in flux state is determined by this prior, where $\gamma = \exp^{-ncp_{prior}}$ represents the constant factor that quantifies the likelihood of detecting k + 1 transition points as opposed to k points (i.e.,  the likelihood of a false positive). It is important to note that the choice of \textit{ncp$_{prior}$} depends on factors such as the length of the LC and the uncertainties associated with the flux measurements. The values of the constant flux amplitude that represent the flux at each block, which are defined by the position of the change points, are obtained calculating the averages of the corresponding daily measurements, weighted by the inverse square of the individual flux uncertainties~\citep{Abeysekara_2017a}.

To address this sensitivity to different data conditions, we performed simulations of source's flux by generating synthetic  LCs spanning the same number of days as in the available data. 

We conducted simulations of LCs by introducing  Poissonian fluctuations based on background maps around the coordinates of the sources of interest. Additionally, we injected a Crab-like point source targeting an average flux of 1~CU. It is worth noting that our simulation approach differs from the one employed in the HAWC monitoring, which was previously used to generate blazar LCs~\citep{Abeysekara_2017a}. In the previous method, a constant flux hypothesis was utilized, following a Gaussian distribution with a standard deviation consistent with the measured uncertainty in the data for the sources of interest.

For our simulations, we set a false positive rate (FPR) of 5\% indicating the probability of detecting at least one additional flux transition beyond the one corresponding to the injected base flux.

We determined the HAWC flux for each BB by treating the normalization $N_{0}$ and the photon index $\alpha$ as free parameters. This approach is feasible due to the significant signal-to-background ratio observed in the majority of the BBs, which allows for constrains on both parameters. 

\section{Results and Discussion}\label{sec:results}
\subsection{Light Curves}\label{sec:LCs}
\begin{figure*}[h!]
\centering
\begin{tabular}{c}
\includegraphics[width=0.80\textwidth]{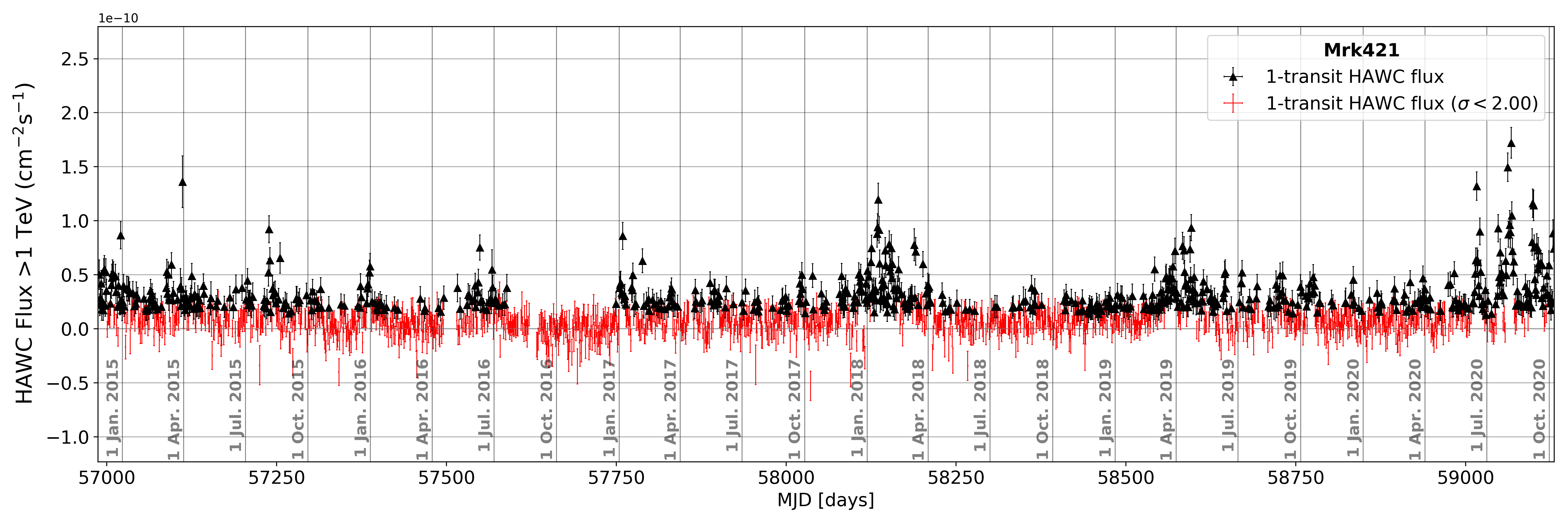}\\
(a)\\\label{fig:mrk421LC}
\includegraphics[width=0.80\textwidth]{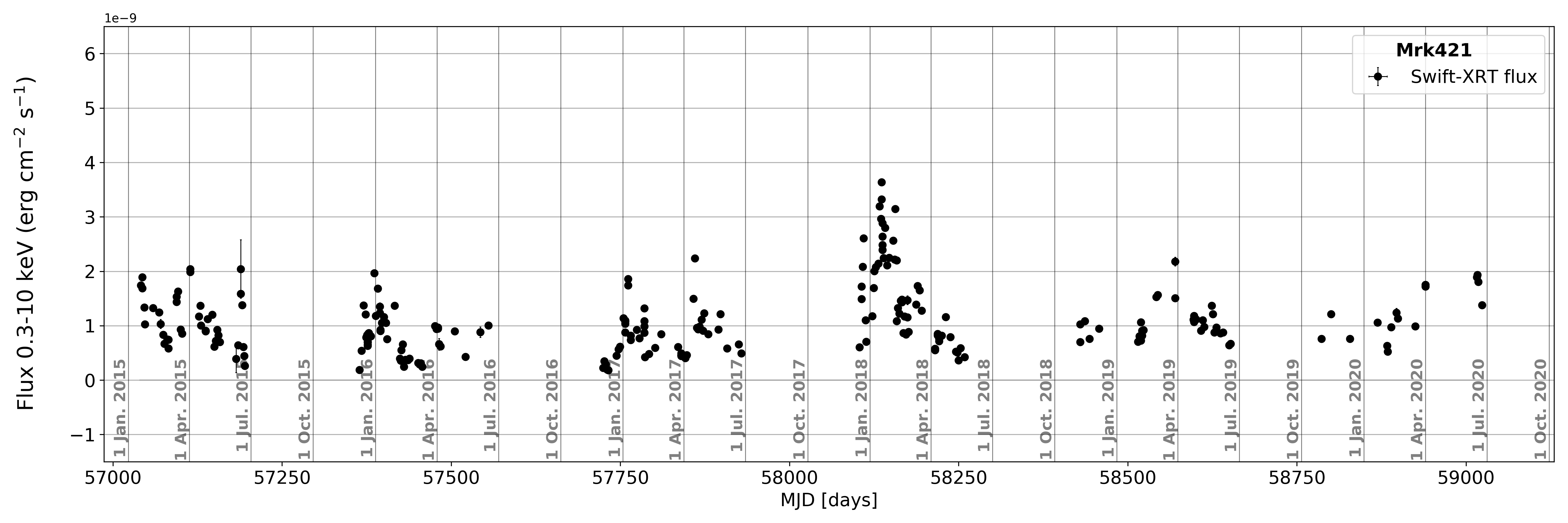}\\
(b)\\\label{fig:xrtmrk421LC}
\includegraphics[width=0.80\textwidth]{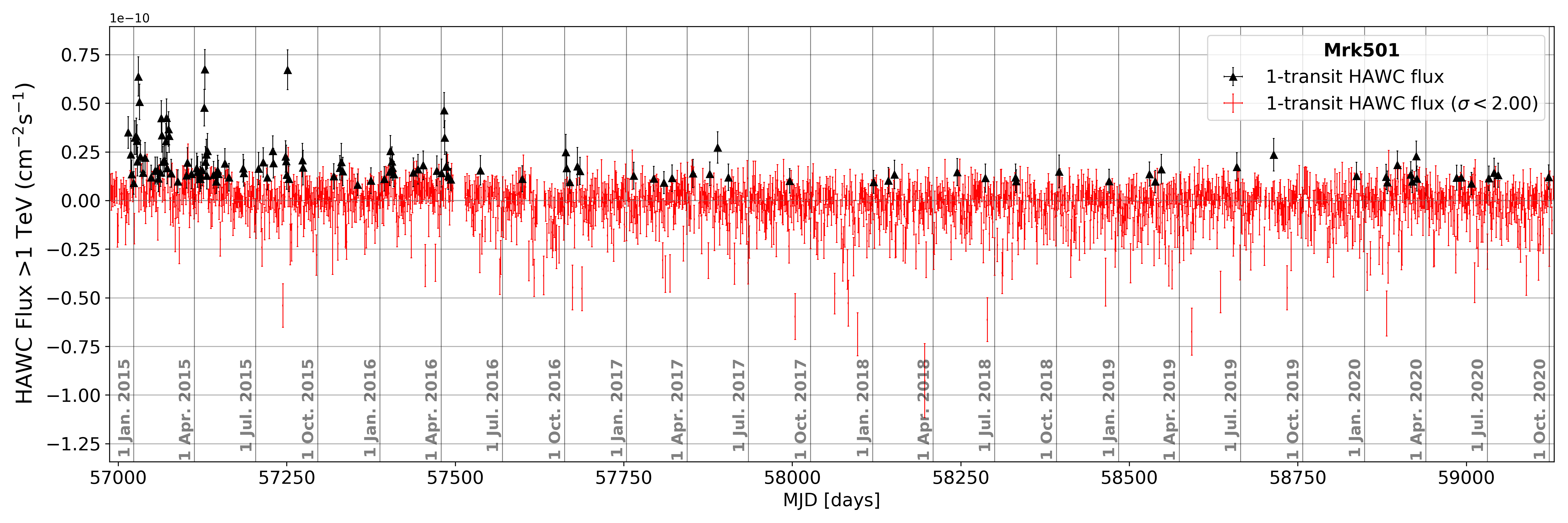}\\
(c)\\\label{fig:mrk501LC}
\includegraphics[width=0.80\textwidth]{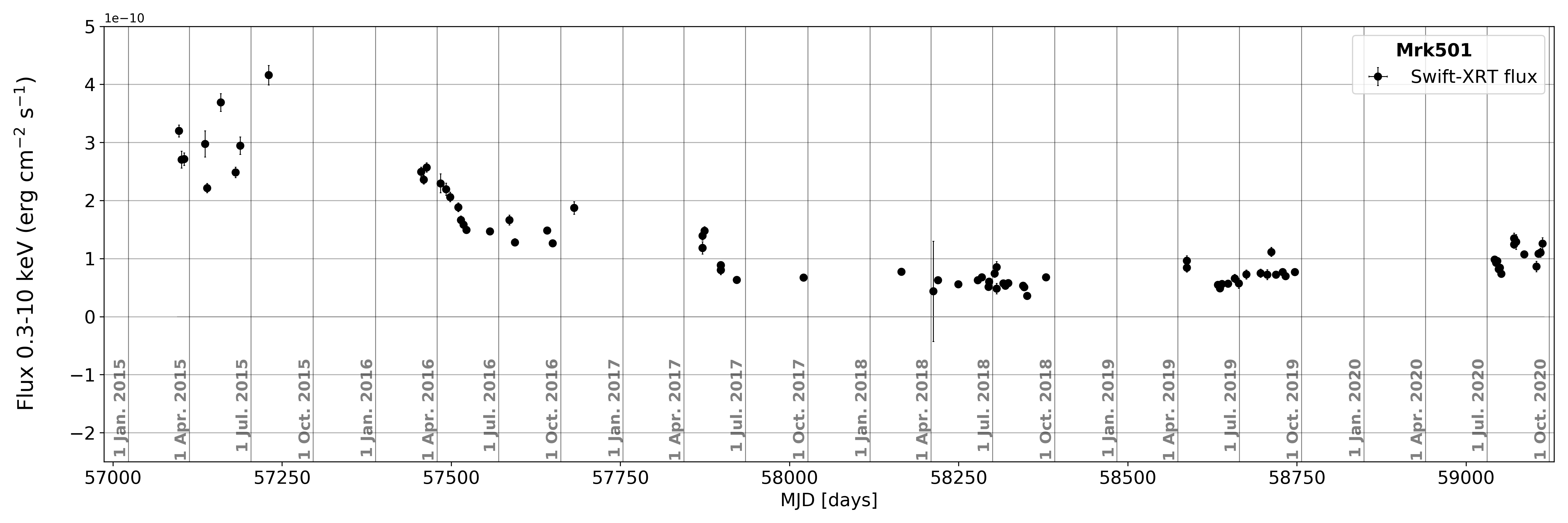}\\
(d)\label{fig:xrtmrk501LC}
\end{tabular}
\caption{ LCs for Mrk~421 and Mrk~501. HAWC LC showing Integrated flux $>1$ TeV, (a) and (c) panels. A red marker is showing when the significance in the flux was below $2\sigma$ for that transit. For the \textit{Swift}-XRT LC, (b) and (d) panels, showing integrated flux between 0.3-10 keV.\\ \label{fig:LCnoBBMrk421}}
\end{figure*}

For Mrk~421 and Mrk~501, we calculated the $\gamma$-ray flux $>1$ TeV for each HAWC transit, which has a duration of approximately 6.2 hr, and the corresponding X-ray flux in the 0.3\textendash10 keV energy band   using the data and software detailed in Section \ref{sec:analysisHAWC} and \ref{sec:analysisXRT} respectively. The resulting $\gamma$-ray and X-ray LCs are presented in Figure~\ref{fig:LCnoBBMrk421}. In this figure, we only display X-ray fluxes that overlap in time with HAWC observations for Mrk~501 and Mrk~421, which amount to \tranfooneCorSize~and \tranftwoneCorSize~ data points respectively. We have highlighted in red all the daily fluxes with a significance below $2\sigma$. In the case of Mrk~501, we have an integrated flux distribution dominated by low significance values, which also leads to negative flux values. We checked the significance distribution for the source and also in the case when there is only a background expectation, and we found them to be consistent to each other. 
The significance distributions are centered at zero and they are not fully Gaussian, showing a broader distribution with an average standard deviation of ~1.5 sigma. This leads to higher statistical fluctuations to more extreme significance values as expected for a Gaussian distribution.

Mrk~501 exhibited low activity during the majority of the observational period, spanning over two years. However, there was an  initial burst of activity between January and June 2015, during which both the $\gamma-$ray and the X-ray fluxes increased by a factor of approximately 5 and 3, respectively. A brief flare is discernible around April 2016, with X-ray and $\gamma$-ray emissions, the former seemingly lasting longer than the latter. 

In contrast, Mrk~421 displayed several periods of heightened activity in both energy bands. The highest X-ray fluxes occurred between January and April 2018, while the highest $\gamma$-ray fluxes, equivalent to $\sim$7 Crab units, were observed towards the end of the HAWC observational period, after June 2020. Notably, Mrk~421 appears to be active simultaneously in both energy bands. It is worth mentioning that during the period after June 2020, the LC exhibits four distinct peaks, each lasting approximately 15 days. A more detailed analysis of these flares is planned for future studies.

\subsection{\texorpdfstring{X-ray/$\gamma$-ray  }~correlation per transit}\label{sec:xraycorrtrans}
To investigate the presence of a correlation between the emissions in both energy bands, we adopt the procedure outlined in Section \ref{sec:xraycorr} and start with the simplest model: a linear correlation. This choice is motivated, particularly for Mrk~421, by previous studies conducted in similar energy bands, for example \citet{2019MNRAS.484.2944G, 2003ApJ...598..242A,Acciari_2020a,2005ApJ...630..130B}.

To calculate the X-ray/$\gamma$-ray correlation, we use the HAWC transits that have a corresponding X-ray flux,  then we are only considering the remaining transits when a minimum of $2\sigma$ for the $\gamma$-ray flux significance was reached.

\begin{figure*}[ht]
\begin{tabular}{cc}
    \includegraphics[width=0.49\textwidth]{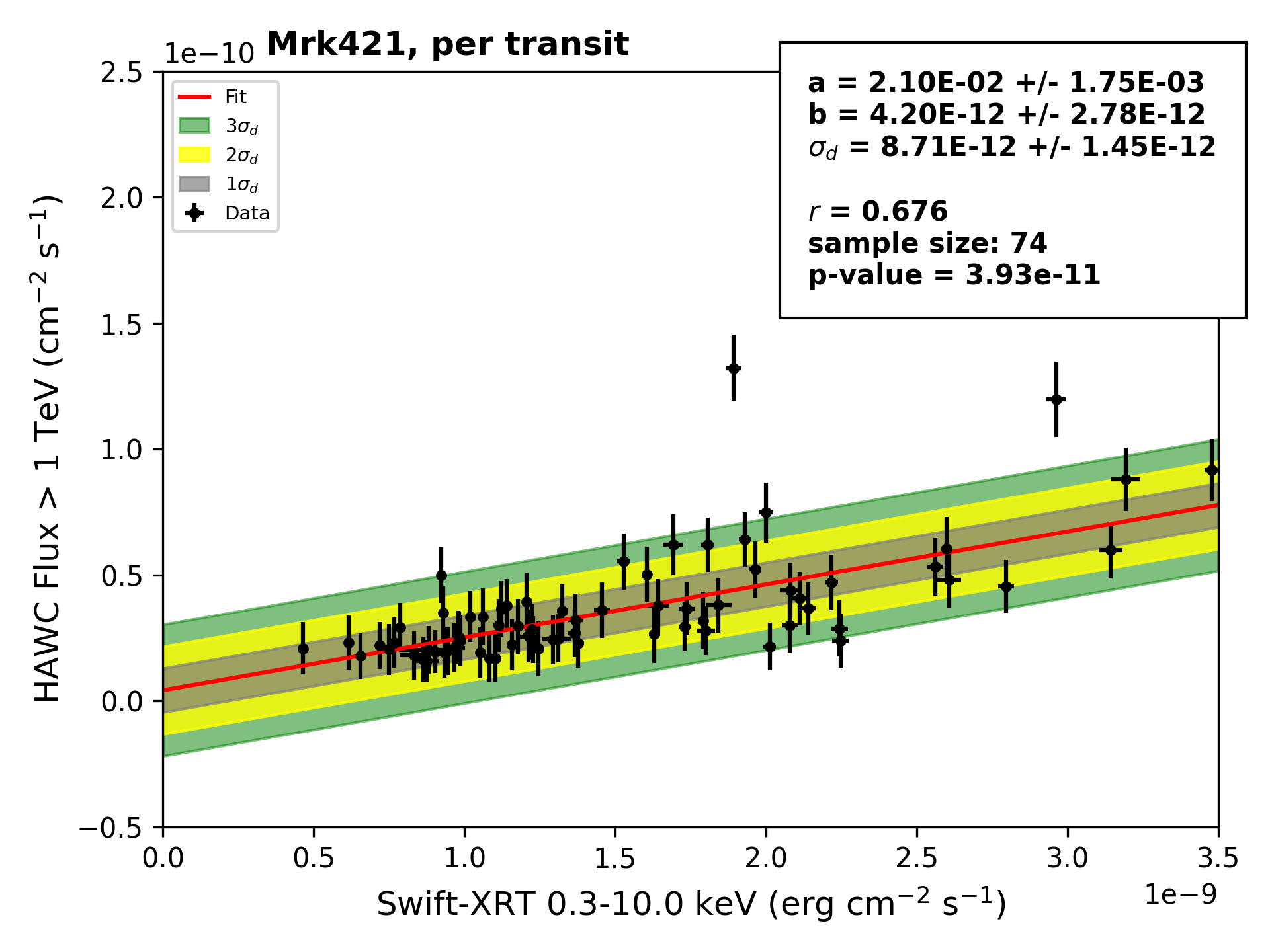}\label{fig:corrplottrans421} &   \includegraphics[width=0.49\textwidth]{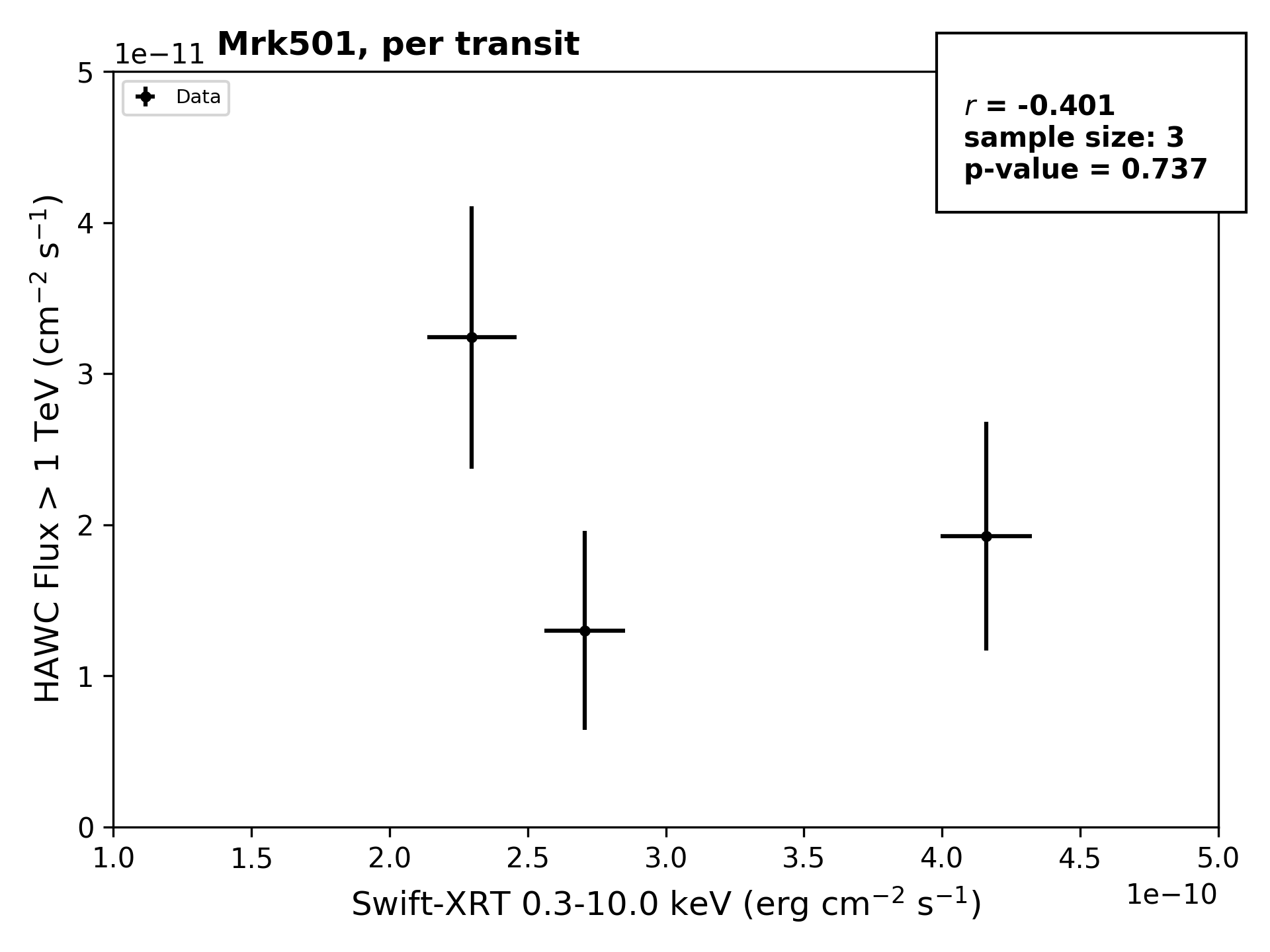}\label{fig:corrplottrans501} \\
(a) & (b) \\[6pt] \\
    \includegraphics[width=0.49\textwidth]{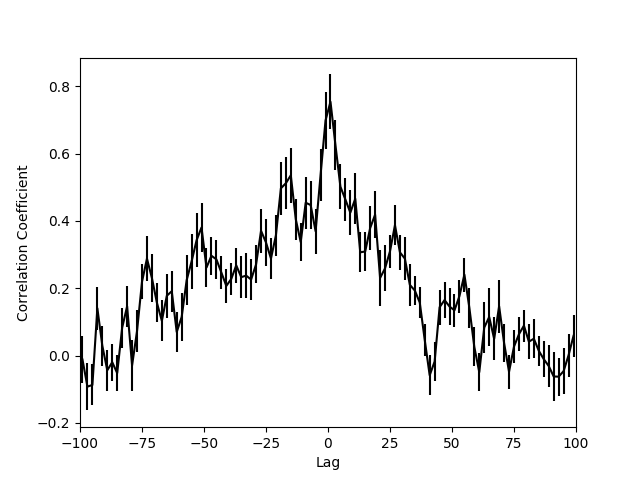}\label{fig:dcf421} &   \includegraphics[width=0.49\textwidth]{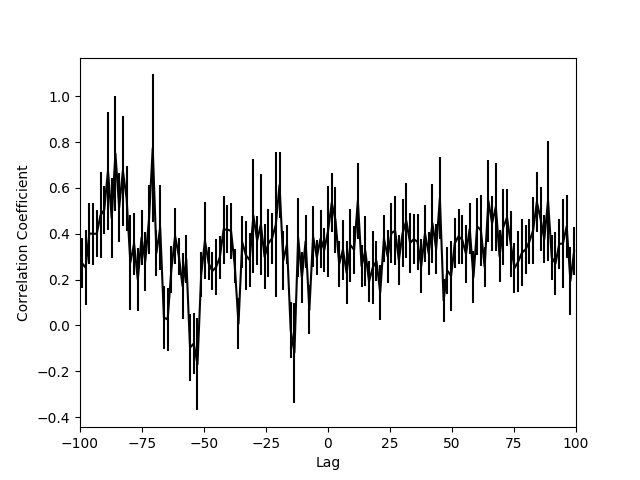}\label{fig:dcf501} \\
    (c) & (d) \\[6pt] \\
\end{tabular}
\caption{X-ray/$\gamma$-ray correlation for Mrk~421 per transit (a). The p-value of the correlation is \tranftwoneCorR~that shows consistency with a linear correlation. X-ray/$\gamma$-ray correlation for Mrk~501 per transit (b). It has been noticed that Mrk~501 has had a long period of very low activity that is reflected in a low flux measured even when higher activity in the X-ray emission happens, resulting in only 3 data points as mentioned in~\ref{sec:xraycorrtrans}. Discrete correlation function (DCF) for Mrk~421 (c). We can see a correlation between the two bands for no time lag. In the case of Mrk~501 (d) there is no evidence of correlation 
}\label{fig:corrplottrans}
\end{figure*}

The results of the correlation for both sources are presented in Figure~\ref{fig:corrplottrans} (panels a,b) and in the case of Mrk~421, the values of the parameters are summarized on table \ref{tab:corrFit} (row 1). For Mrk~501 we only obtained 3 data points, therefore we can not obtain a reliable measurement for the correlation per transit.   

We also search for cross correlation between the two bands to investigate weather a time lag between the $\gamma$-rays and X-rays was present. We used the discrete correlation function~\citep{1988ApJ...333..646E} as implemented in~\citet{10.1093/mnras/stv1575}. The result for Mrk~421 is shown in Fig~\ref{fig:corrplottrans} (c), we obtain a correlation of \DCFMrktwone with no time lag. A similar study~\citep{Acciari_2020b} that used multi-wavelength data from a campaign spanning from 2015  to 2016 also found a strong correlation with no time lag between VHE $\gamma$-rays and X-rays using data from MAGIC (two energy bands, 0.2–1 TeV and $>1$ TeV) and FACT ($E_{th} \sim 0.7$ TeV) for the VHE emission and data from \textit{Swift}-XRT (0.3-10 keV) for the x-ray emission. Additionally, in our work for Mrk~501 (Fig.~\ref{fig:corrplottrans} (d)) there is no evidence of correlation.

\begin{table*}[ht]
    \centering
    \scriptsize
    \begin{tabular}{|>{\centering}p{0.06\textwidth}|>{\centering}p{0.18\textwidth}|>{\centering}p{0.20\textwidth}|>{\centering}p{0.18\textwidth}|>{\centering}p{0.05\textwidth}|>{\centering}p{0.05\textwidth}|>{\centering\arraybackslash}p{0.12\textwidth}|}
    \hline
    Source & a ($\times10^{-2}$) & b ($\rm \times10^{-12} cm^{-2}s^{-1}$) & $\sigma_{d}$ ($\rm \times10^{-12} cm^{-2}s^{-1}$) & $r$ & sample size & p-value \\
    \hline
    Mrk~421 & \tranftwoneCorPa & \tranftwoneCorPb & \tranftwoneCorPs & \tranftwoneCorR & \tranftwoneCorSize & \tranftwoneCorPval \\
    \hline
    Mrk~421 & \bbftwoneCorPa & \bbftwoneCorPb & \bbftwoneCorPs & \bbftwoneCorR & \bbftwoneCorSize & \bbftwoneCorPval \\
    Mrk~501 & \bbfooneCorPa & \bbfooneCorPb & \bbfooneCorPs & \bbfooneCorR & \bbfooneCorSize & \bbfooneCorPval \\
    \hline
    Mrk~421 & \SIbbftwoneCorPa & \SIbbftwoneCorPb & \SIbbftwoneCorPs & \SIbbftwoneCorR & \SIbbftwoneCorSize & \SIbbftwoneCorPval \\
    \hline
    \end{tabular}
    \caption{Correlation fit parameters for both Mrk~421 and Mrk~501 using \dagostini{ }~\citep{dagostini1} approach in Equation~\ref{eq:dagostinie} to estimate the robustness of the linear fit by introducing an extra scatter parameter $\sigma_{d}$. Data used in the fit corresponds to quasi-simultaneous observations of HAWC and \textit{Swift}-XRT. The binning used is 1 HAWC transit (line 1) and BBs (lines 2 and 3). The sample size is the number of sidereal days or BBs that overlap for both data samples. Line 4 corresponds to the Flux vs. photon index correlation for Mrk~421 (Figure~\ref{fig:corrBBplotGindex})}
    \label{tab:corrFit}
\end{table*}{}


In the case of Mrk~421, there are a few outliers where the highest $\gamma$-ray 
fluxes are much higher than the values predicted by the correlation for that given value of X-ray flux. Surprisingly, this does not seem to significantly affect the PCC value, considering the obtained p-value and the size of the sample. However, even with the introduction of $\sigma_{d}$, it does not fully account for the dispersion observed beyond the correlation of these outliers. These results do not provide a clear explanation for whether the correlation breaks down for $\gamma$-ray fluxes beyond a value around $\rm 1\times10^{-10}\  cm^{-2}s^{-1}$ for energies greater than 1 TeV, or if there are specific activity states where either the X-ray flux is suppressed or contributions from other processes affect the $\gamma$-ray flux. It is also possible that these observations result from the quasi-simultaneity of the data and the high variability of the source. HAWC fluxes are measured over extended periods of hours, whereas \textit{Swift} observations coincide with  HAWC but last on the order of 1 ks. If this were the case, we should expect to observe high X-ray fluxes alongside $\gamma$-ray fluxes that are lower than expected. However, this is not observed in at least this data set.

\subsection{\texorpdfstring{X-ray/$\gamma$-ray  }~correlation in BBs}\label{sec:xraycorrBB}
The large fluctuations in the data around the correlation may have various origins. These fluctuations could result from observational statistical uncertainties, be intrinsic to the emission mechanisms, or be influenced by the quasi-simultaneity of the data. To investigate the origin of the fluctuations around the correlation, we employ Bayesian Blocks. The BBs in each HAWC data set are obtained by following the procedure described in section \ref{sec:bb}. The value of the $ncp_{prior}$ parameter was set to \ncpused~ for both Mrk~501 and Mrk~421, corresponding to a FPR of $5\%$.

\begin{figure*}[ht]

\gridline{\fig{mrk501_maps_a_Nov26-2014-Oct07-2020_ncp_9.5_nodata.png}{0.98\textwidth}{(b)\label{fig:mrk421LCBB}} }

\gridline{\fig{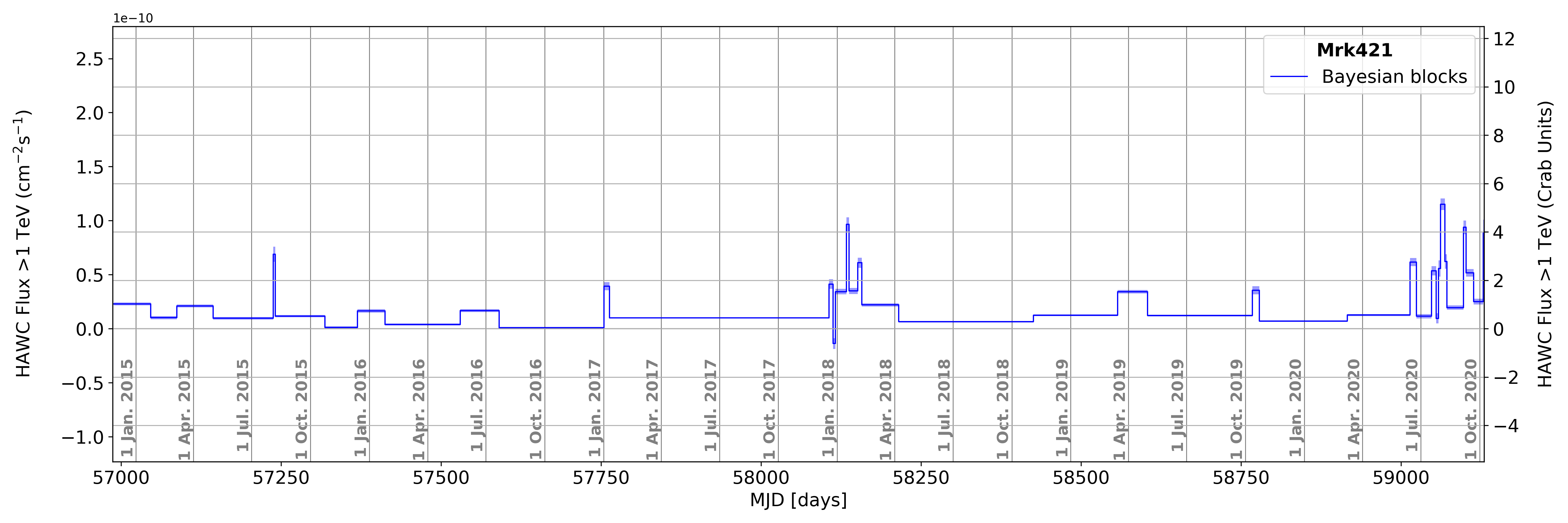}{0.98\textwidth}{(a)\label{fig:mrk501LCBB}}
          }   
\caption{HAWC LCs for (a) Mrk~501 (b) Mrk~421 for an Integrated flux $>1$TeV within the source transit ($\sim6.2$ hr) using a total of \daysused\  after applying the BB algorithm.\label{fig:BBLCs}}
\end{figure*}

We only obtained 5 BBs that have overlap with X-ray data, leading to a low statistics that will not be useful to study the correlation, although we did calculate it for completeness. However, in the case of Mrk~421, we identify twice the number of BBs compared to those reported by \citet{Abeysekara_2017a}. This increase is expected, given the extended duration of the analyzed data set, although the increase is probably not proportional to the amount of data added. However, this can be explained by the moderate and variable fluxes before April 2017 and for the only three distinct phases of higher activity, each of which was preceded by a period of low activity. In both LCs, longer BBs tend to absorb outliers due to the short-lived nature of emissions with the highest fluxes. Detailed information regarding the start and end time of these BBs, for both Mrk~501 and Mrk~421 as well as the corresponding integrated $\gamma$-ray flux values $> 1$ TeV is provided in Table~\ref{tab:mrk501bbt}. We are only showing BBs when the integrated flux significance was $> 2$ sigma. Additionally, Figure \ref{fig:BBLCs} displays the BBs for both sources. 

\begin{table*}[ht]
    \centering
    \scriptsize
    \begin{tabular}{c|c|c|c}
    \hline
    MJD Start & MJD Stop & Duration & Flux $> 1$ TeV\\
    &  & (days) & $(\rm cm^{-2}s^{-1} \times 10^{-11})$\\
    \hline
    \multicolumn{4}{c}{Mrk~501}\\
    \hline
57076.42 & 57126.54  &  50.86   &  $0.82 \pm 0.22$ \\
57133.27 & 57250.20  &  117.68  &  $0.64 \pm 0.11$ \\
57251.94 & 57565.34  &  314.14  &  $0.56 \pm 0.09$ \\
57568.08 & 58093.89  &  528.55  &  $0.15 \pm 0.03$ \\
58097.63 & 59129.06  &  1031.18 &  $0.07 \pm 0.07$ \\
    \hline
\multicolumn{4}{c}{Mrk~421}\\
\hline
56988.42  & 57045.52  & 57.842   & $2.55  \pm 0.14 $\\
57046.26  & 57086.40  & 40.888   & $1.17  \pm 0.14 $\\
57087.15  & 57143.25  & 56.844   & $2.45  \pm 0.14 $\\
57144.00  & 57236.99  & 93.743   & $1.18  \pm 0.11 $\\
57237.74  & 57239.99  & 2.992    & $7.07  \pm 0.71 $\\
57240.73  & 57317.77  & 77.787   & $1.37  \pm 0.11 $\\
57318.52  & 57368.63  & 50.861   & $0.21  \pm 0.11 $\\
57369.38  & 57411.51  & 42.883   & $1.79  \pm 0.15 $\\
57412.26  & 57529.19  & 117.678  & $0.49  \pm 0.09 $\\
57529.94  & 57590.03  & 60.833   & $1.87  \pm 0.14 $\\
57590.77  & 57753.58  & 163.552  & $0.17  \pm 0.07 $\\
57754.33  & 57762.55  & 8.975    & $4.12  \pm 0.38 $\\
57763.30  & 58105.61  & 343.061  & $1.19  \pm 0.05 $\\
58106.36  & 58111.60  & 5.984    & $4.21  \pm 0.44 $\\
58116.33  & 58132.54  & 16.954   & $3.28  \pm 0.25 $\\
58133.29  & 58136.53  & 3.989    & $9.73  \pm 0.65 $\\
58137.28  & 58150.49  & 13.962   & $3.72  \pm 0.29 $\\
58151.24  & 58156.48  & 5.984    & $6.20  \pm 0.48 $\\
58157.22  & 58212.32  & 57.842   & $2.51  \pm 0.15 $\\
58215.07  & 58424.74  & 210.424  & $0.76  \pm 0.07 $\\
58425.49  & 58556.38  & 131.640  & $1.36  \pm 0.08 $\\
58557.13  & 58603.25  & 46.872   & $3.67  \pm 0.16 $\\
58604.00  & 58766.81  & 163.552  & $1.52  \pm 0.08 $\\
58767.55  & 58777.78  & 10.970   & $4.20  \pm 0.43 $\\
58778.52  & 58915.40  & 137.623  & $0.85  \pm 0.07 $\\
58916.14  & 59013.13  & 97.732   & $1.49  \pm 0.10 $\\
59013.88  & 59022.11  & 9.973    & $6.45  \pm 0.38 $\\
59023.85  & 59047.04  & 23.934   & $1.50  \pm 0.22 $\\
59047.79  & 59054.02  & 6.981    & $5.74  \pm 0.43 $\\
59054.77  & 59058.01  & 3.989    & $1.19  \pm 0.46 $\\
59058.75  & 59060.00  & 2.992    & $5.53  \pm 0.75 $\\
59061.75  & 59067.98  & 6.981    & $10.91 \pm 0.50 $\\
59068.73  & 59070.97  & 2.992    & $6.35  \pm 0.67 $\\
59071.72  & 59096.90  & 25.929   & $2.13  \pm 0.22 $\\
59097.65  & 59100.89  & 3.989    & $9.78  \pm 0.64 $\\
59101.64  & 59112.86  & 11.967   & $5.99  \pm 0.37 $\\
59113.60  & 59127.82  & 14.959   & $2.72  \pm 0.27 $\\
59128.56  & 59129.81  & 0.997    & $8.55  \pm 0.85 $\\
    \hline
    \end{tabular}
    \caption{HAWC fluxes for  \mrkbbfoone~and \mrkbbftwone~found in the LCs of Mrk~501 and Mrk~421 respectively, when utilizing \daysused~of data and a FPR of 5\% was chosen.}
    \label{tab:mrk501bbt}
\end{table*}

\begin{figure*}[ht]
\begin{tabular}{cc}
    \includegraphics[width=0.49\textwidth]{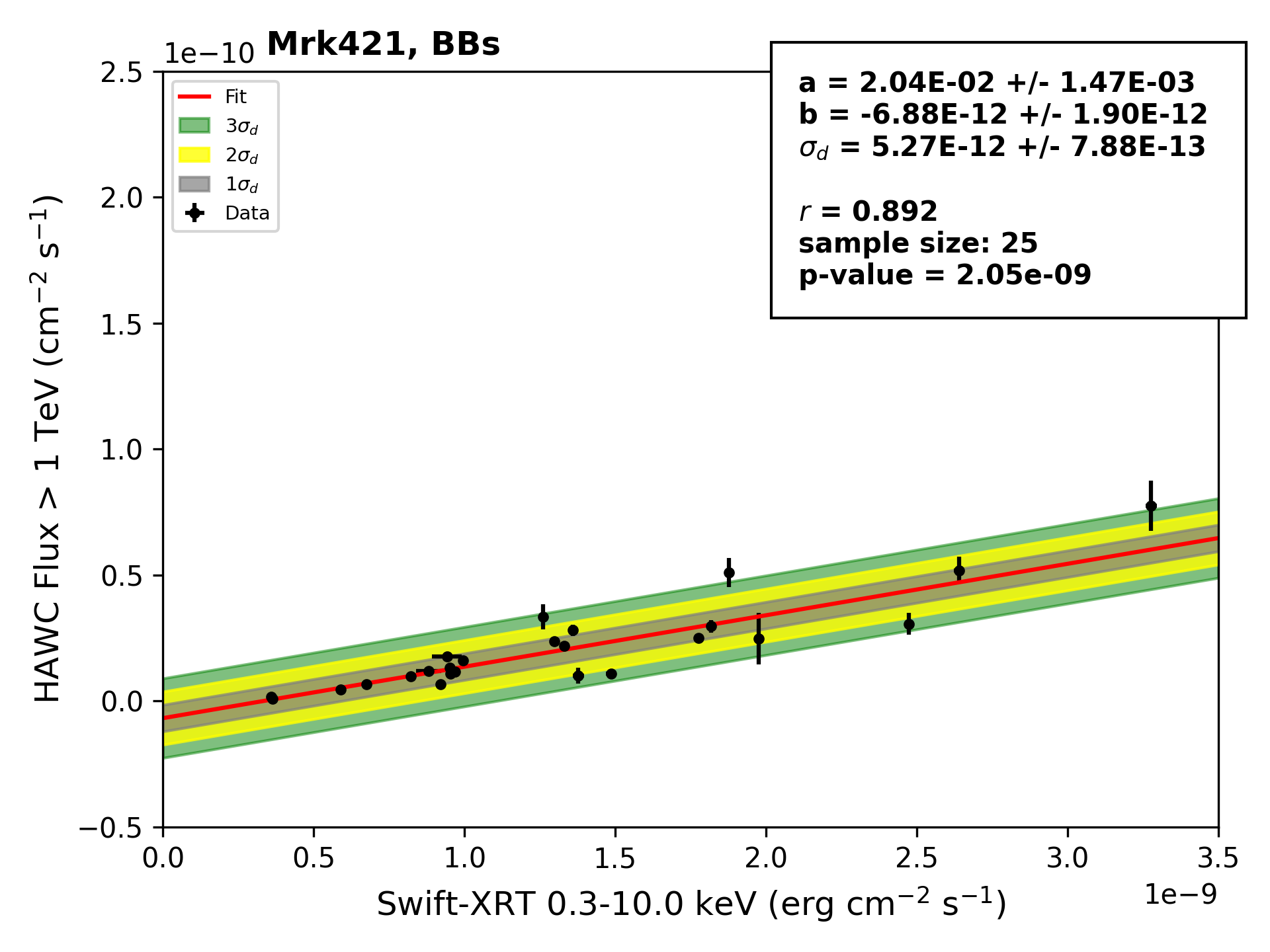} &   \includegraphics[width=0.49\textwidth]{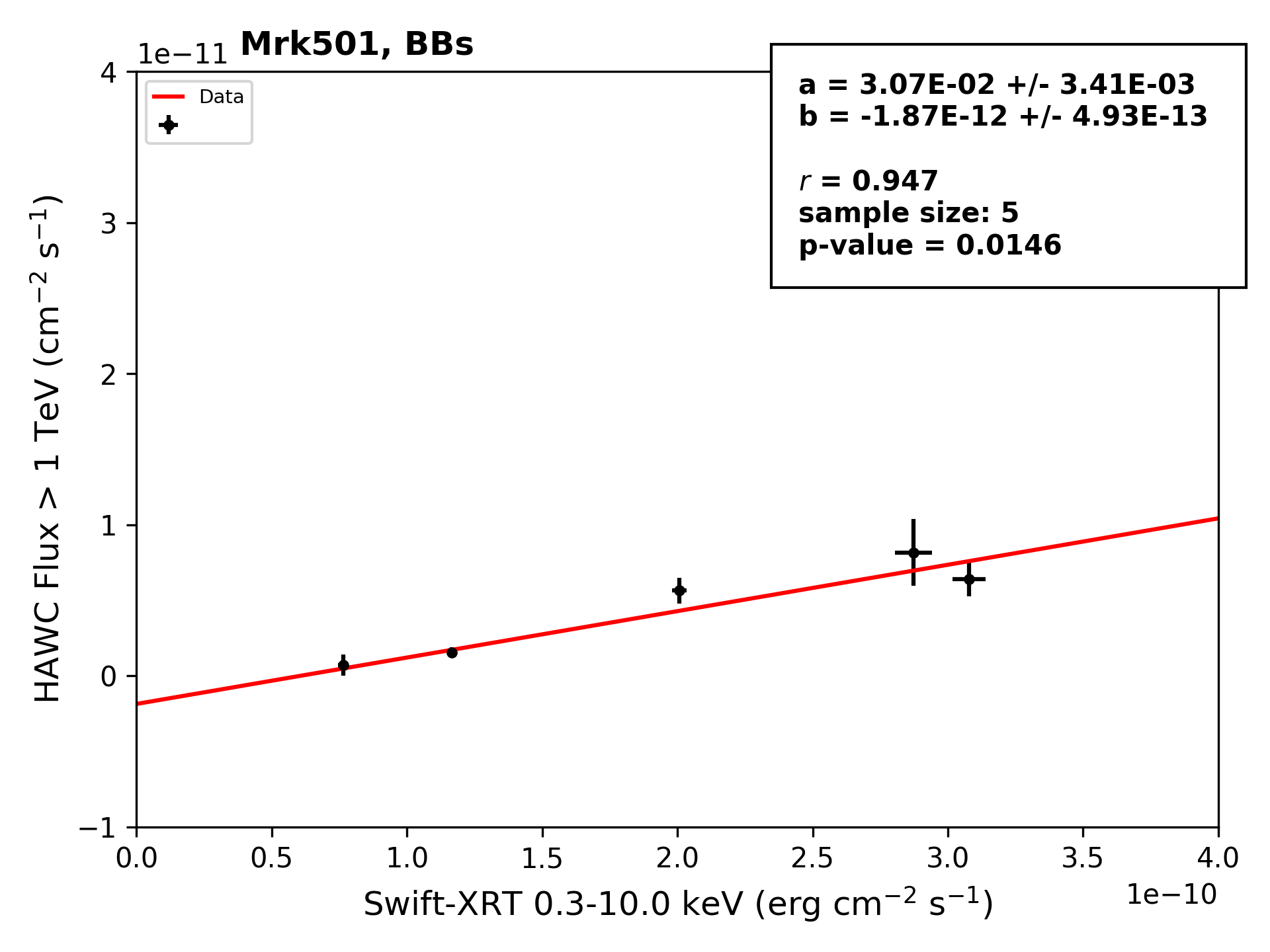} \\
(a) & (b) \\[6pt]
\end{tabular}
\caption{X-ray/$\gamma$-ray correlation for Mrk~421 in BBs (a). The p-value of the correlation is \bbftwoneCorR~that shows a strong consistency with a linear correlation. X-ray/$\gamma$-ray correlation for Mrk~501 in BBs (b). Due to the small size of the sample and the p-value, we only have evidence of a linear correlation.}\label{fig:corrplottrans2}
\end{figure*}

Using the BBs, we once again follow the procedure outlined in Section \ref{sec:xraycorr} to examine a linear correlation. The flux values reported in Table~\ref{tab:mrk501bbt} are the integrated flux for the data within the BB and both the flux normalization and the spectral index were set as free parameters. The results of this analysis are presented in Figure \ref{fig:corrplottrans2}. Notably, the outliers are incorporated into the correlation through BBs with lower fluxes. As a consequence, the slopes of the correlations become less steep but remain consistent with the previously obtained values within twice their statistical uncertainties. Additionally, the values of $\rm b$ approach zero, although they still deviate from a zero value.
\subsection{Correlation Interpretation}\label{sec:discussion}
In the case of Mrk421, the intrinsic dispersion ($\sigma_{d}$) is slightly smaller than what was obtained using daily fluxes, but it remains significant and is of the same order of magnitude. Interestingly, the data dispersion increases as the fluxes increase. Specifically, the correlation is tighter for X-ray fluxes lower than 1.25 $\rm erg\ cm^{-2} s^{-1}$, while for fluxes higher than 1.75 $\rm erg\ cm^{-2} s^{-1}$, most of the data points deviate up to 3$\sigma_{d}$ from the correlation. This could be indicative that the fluctuations are intrinsic to the source. However, this phenomenon is not observed in Mrk~501, possibly due to the lack of measured high-flux states in the data sample used.

According to the SSC model, for a single flare period, the expected correlation between X-ray and $\gamma$-ray fluxes is quadratic, and a change in particle density is responsible for the increase in the $\gamma$-ray flux \citep{refId0_2005, refId0-2010-2, 1999ApJ...514..138K}. However, it has been demonstrated by several authors that this is not the case for Mrk~421 \citep{2019MNRAS.484.2944G, 2003ApJ...598..242A, Acciari_2020a, 2005ApJ...630..130B}.

Mrk~421, in most observations, exhibits a linear correlation, and this linearity is generally attributed to contributions from multiple emission zones \cite{Katarzyski2010}, external inverse Compton with seed photons from the accretion disk \citep{dermer_1992}, or emission from the central region of the blazar \citep{sikora_1994}. It may also result from emission in the Klein-Nishina energy regime \citep{katarzynski2005correlation}. For Mrk 421, depending on the model and activity state, the contribution to the TeV flux comes from photon scattering by the highest-energy electrons in the Klein-Nishina regime \citep{2005A&A...437...95A}. Furthermore, most short-term TeV observations of Mrk 421 are consistent with a one-zone scenario, as reported by~\citet{2011ApJ...736..131A,2021A&A...655A..93A,10.1093/mnras/stac917}, suggesting that there is no need for external photons.

On the contrary, it is evident that the HAWC data set presented here for Mrk~421 does not originate from a single flaring event; rather, it includes at least four well-separated flaring events during the long-term monitoring data spanning 6 years. As noted in \citep{2005A&A...437...95A}, the long-term correlation is expected to differ from that of a single flaring event due to the inclusion of multiple flaring events, each driven by different processes that increase the TeV emission. These processes can include changes in particle densities or magnetic fields. 
The range of theoretical parameters describing flaring episodes at different multi-wavelength campaigns suggests that multiple emission zones may be involved \citep{2013ApJ...768...54B,2015A&A...578A..22A,2015MNRAS.448.3121H, 2019ApJ...886...23X, 2021MNRAS.505.2712D, 2023MNRAS.526.5054L}. Consequently, the long-term correlation can be seen as analogous to the correlation expected from multiple emission zones. Therefore, a linear correlation would be expected, as obtained in this work.

If we consider $F_{\gamma} \propto A F_{X}^{\beta}$, the total normalization of the correlation, represented by $A$, for a single flaring event, depends strongly on the initial size of the blob, the initial particle density for electrons with energy above the break energy, and the initial magnetic field, as discussed in \citet{2005A&A...437...95A}. Therefore, assuming that these values vary slightly from flare to flare while remaining within the average values for a given blazar, we could expect the dispersion of the correlation to result from variations in these parameters.
\subsection{\texorpdfstring{$\gamma$-ray }~flux vs. photon index}\label{sec:gammaVsindex}
Previous analyses have reported a ``harder-when-brighter'' behavior in the relationship between spectral indices and their corresponding $X$-ray fluxes \citep{corrindex}, as well as in the relationship between the $\gamma$-fluxes and their corresponding spectral indices~\citep{Acciari_2020b}.
Since the synchrotron and SSC components are produced by the same electron population, we can expect their energy dependencies to be similar~\citet{corrindex}.

\begin{figure*}[h!]
\centering
\begin{tabular}{cc}
    \includegraphics[width=0.47\linewidth]{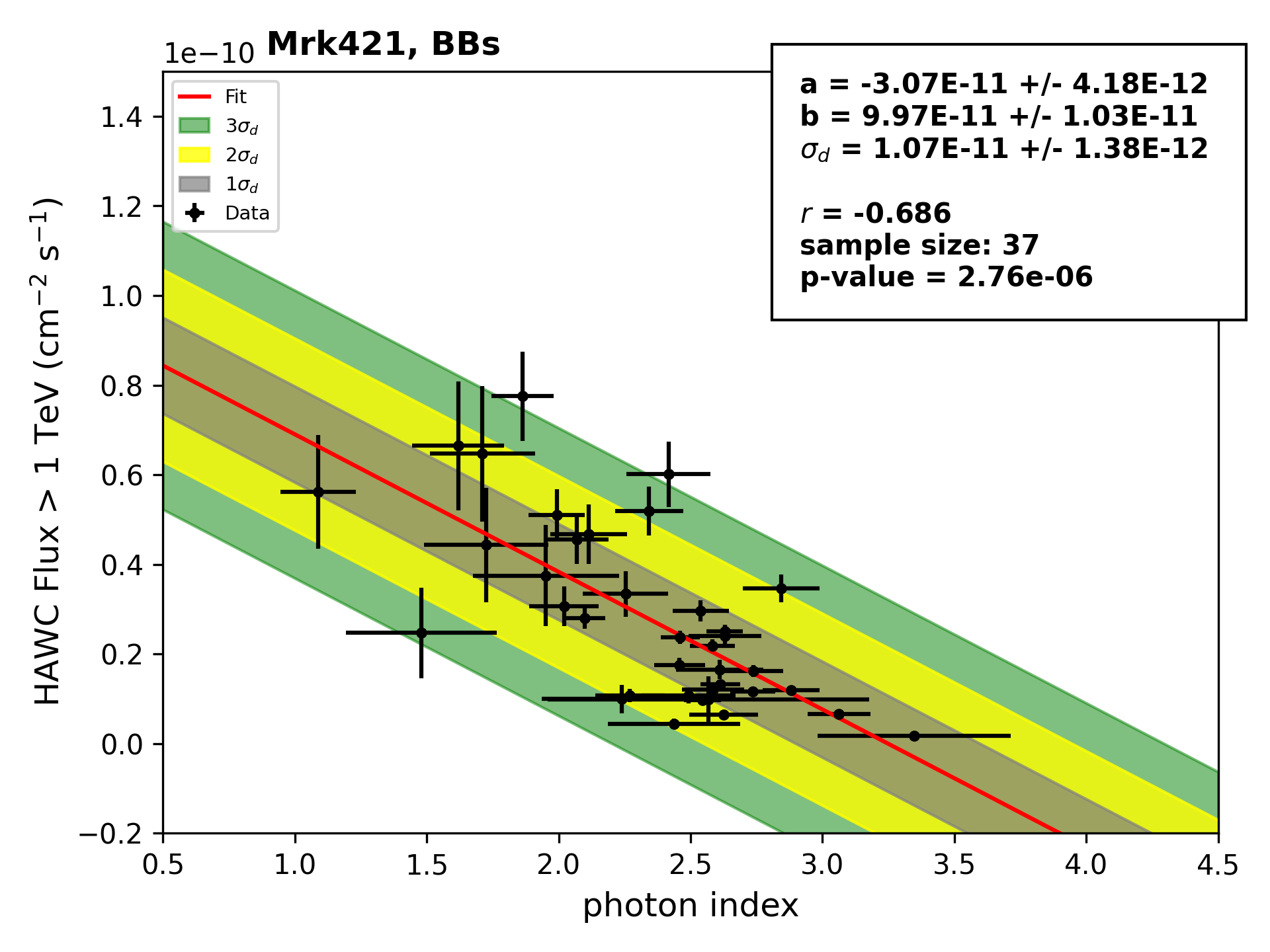} & \includegraphics[width=0.47\textwidth]{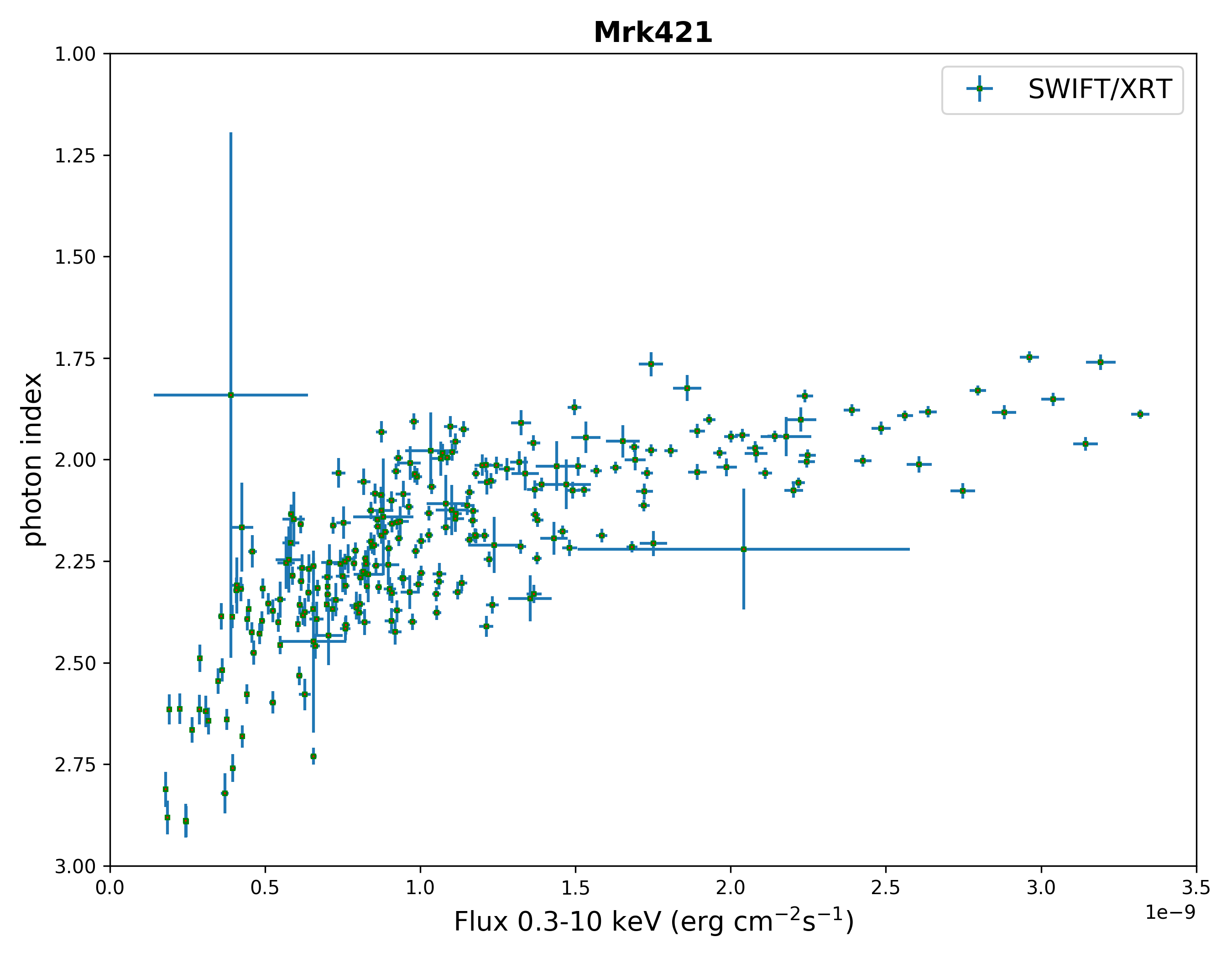} \\
    (a) & (b)\\[6pt]
    \includegraphics[width=0.47\textwidth]{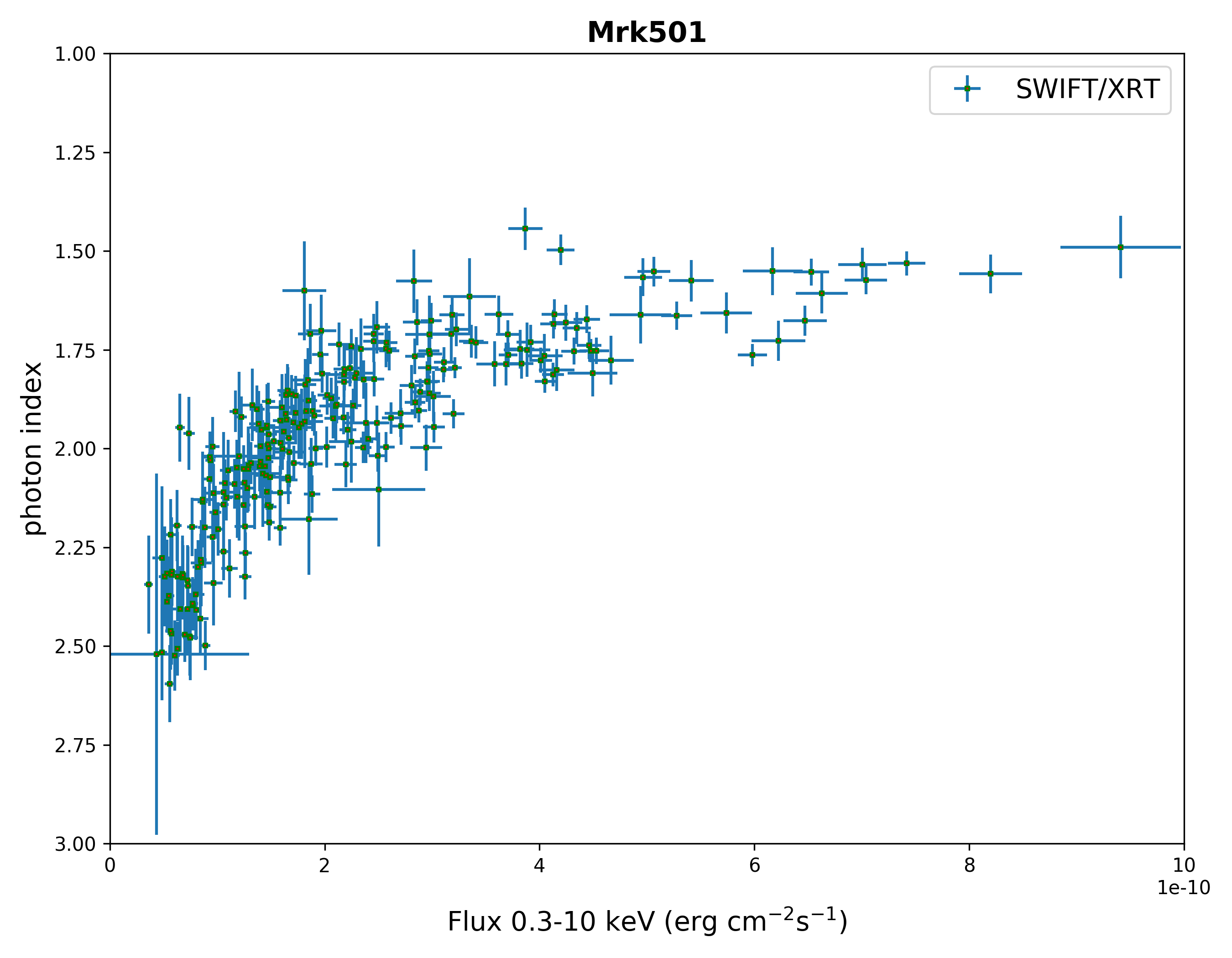} & \\
    (c) & \\[6pt]
\end{tabular}
\caption{$\gamma$-ray flux vs. photon index (a) and photon index vs. X-ray flux (b), (c). We observed a ``harder-when-brighter'' behavior for Mrk~421, revealing a linear relationship within the energy bands employed. However, in the case of the X-ray flux shown in (b), a linear correlation with the photon index is absent, Nevertheless, the ``harder-when-brighter'' trend observed in the $\gamma$-ray flux persists. Notably, X-ray flux values below 1.1 $\times 10^{-9} \rm erg\ cm^{-2} s^{-1}$ correspond to $\gamma$-ray fluxes closer to the linear correlation displayed in Fig.~\ref{fig:corrplottrans}(a). For Mrk~501, a similar behaviour to Mrk~421 is seeing between the X-ray flux and the photon index, as depicted in (c). However, less dispersion is observed in the data for X-ray flux below 3 $\times 10^{-10} \rm erg\ cm^{-2} s^{-1}$.}\label{fig:corrBBplotGindex}
\end{figure*}

We observed the same ``harder-when-brighter'' behavior for Mrk~421, as depicted in Figure \ref{fig:corrBBplotGindex}, panel a. We employed the maximum likelihood method, as discussed in Section \ref{sec:xraycorr}, to establish the optimal linear representation of $\gamma$-ray fluxes observed by HAWC as a function of the corresponding spectral indices for the BBs displayed in Figure \ref{fig:BBLCs}. We also depict the photon indices for the X-ray emission as function of the corresponding fluxes for Mrk~421 (panel b) and Mrk~501 (panel c). Although we observe a ``harder-when-brighter'' behavior on the X-ray fluxes for Mrk~421, it is far from being linear, consistent with the findings of \citet{corrindex}.

In this work, we find that the $\gamma$-ray flux for Mrk~421 is negatively correlated to its photon index, with a Pearson correlation coefficient (PCC) of \SIbbftwoneCorR, and a slope of \SIbbftwoneCorPa. Unfortunately, no conclusion for Mrk~501 can be drawn due to its low activity during the studied period.

Two noteworthy points should be emphasized. First, the X-ray photon index remains relatively stable once the X-ray flux reaches approximately 1.1 $\times 10^{-9} \rm erg\ cm^{-2} s^{-1}$ and 3 $\times 10^{-10} \rm erg\ cm^{-2} s^{-1}$ for Mrk~421 and Mrk~501, respectively.

Secondly, in the case of Mrk~421, we observed a significant increase in the data dispersion around the best fit correlation line. These deviations occur beyond the X-ray flux threshold of $1.1 \times 10^{-9} \rm erg\ cm^{-2} s^{-1}$, with the emergence of outliers when using daily fluxes Figure~\ref{fig:corrplottrans} (a)) and when employing BBs (Figure~\ref{fig:corrplottrans2} (a)). However, such behavior is not observed for Mrk~501, where all X-ray fluxes remain below $3 \times 10^{-10}  \rm erg\ cm^{-2} s^{-1}$ (Figure~\ref{fig:corrplottrans2} (b)).

\section{Conclusions}\label{sec:conclusions}

We investigated potential correlations between X-ray and $\gamma$-ray fluxes during a long-term monitoring study of Mrk~421 and Mrk~501, utilizing data from the HAWC Observatory. These two sources stand out as the brightest blazars observed by HAWC. Our dataset included daily flux measurements spanning \daysused~days, which is four times longer than the dataset employed in the study by \citet{Abeysekara_2017a}. Instead of relying on \textit{Swift}-BAT data for X-ray emissions, we opted for \textit{Swift}-XRT data. This choice was motivated by the fact that \textit{Swift}-XRT data covers an energy range closely aligned with the position of the synchrotron peak, similar to how the TeV-energy range corresponds to the SSC peak for these two sources. 

The HAWC light curves show that Mrk~501 maintained an extended low state for over two years, as already reported by~\citet{2023ApJS..266...37A}, while Mrk~421 exhibited moderate variability throughout the six-year monitoring period, observed in both X-Ray and $\gamma$-ray data. The most intense flares in each energy band occurred during distinct time periods, with a gap of more than two years between them. However, it is worth nothing that the source displayed simultaneous activity in both energy bands. 

The daily flux data points reveal a long-term correlation for Mrk~421 for quasi-simultaneous observations, this correlation is well described by a linear relationship. For Mrk~501 we could not measure a correlation due to the low statistics obtained.

In Mrk~421, there exist outliers in the correlation, a finding in line with previous observations by \citet{2019MNRAS.484.2944G}. However, the present study does not provide a definitive explanation for whether the correlation breaks down for $\gamma$-ray fluxes beyond a specific threshold, or if other factors, such as the suppression of X-ray flux contributions from additional processes, influence $\gamma$-ray emissions. To delve into the origin of the dispersion in the correlation, we conducted an analysis employing Bayesian Blocks, as defined by the HAWC data. The results obtained indicate a moderate correlation for Mrk~501 and, in the case of Mrk~421, a strong correlation was obtained, similar to the case of daily fluxes. Furthermore, these results imply that the dispersion observed about the line of best-fit correlation are intrinsic to the nature of these sources.

Synchrotron Self-Compton (SSC) models predict a quadratic correlation between X-ray and TeV emissions for individual flares. In these models, X-ray emission is attributed to synchrotron radiation from highly relativistic electrons within the jet, while TeV emission results from inverse-Compton scattering of synchrotron photons \citep{2011ApJ...736..131A}. The presence of a linear correlation may be viewed as the collective effect of multiple emission zones, likely influenced by different processes throughout various flaring events within the time-span considered in this study. Within this context, fluctuations observed around the correlation may be ascribed to slight deviations from the average values of specific parameters, including the initial size of the emission region, electron density above the break energy, and magnetic field strength.

A ``harder-when-brighter'' behavior is observed in the spectra of Mrk~421 within the two energy bands under consideration. Notably, the $\gamma$-ray flux exhibits a linear dependence on the photon index for Mrk~421, while the X-ray flux does not show a linear relationship  with the corresponding photon index. As the photon index in the X-ray emission approaches its hardest value, the correlation appears to weaken. It is worth mentioning that the hardest photon index in the X-ray regime is not the same for the two sources.

This study reaffirms that the overall correlations for Mrk~421 and Mrk~501 are indeed linear, lending support to the leptonic model that involves multiple emission zones. Based on these findings, we can confidently assert that the \emph{harder-when-brighter} behavior is firmly established in the $\gamma$-ray emission of Mrk~421.

\begin{acknowledgements}
We acknowledge the support from: the US National Science Foundation (NSF); the US Department of Energy Office of High-Energy Physics; the Laboratory Directed Research and Development (LDRD) program of Los Alamos National Laboratory; Consejo Nacional de Ciencia y Tecnolog\'{i}a (CONACyT), M\'{e}xico, grants LNC-2023-117, 271051, 232656, 260378, 179588, 254964, 258865, 243290, 132197, A1-S-46288, A1-S-22784, CF-2023-I-645, c\'{a}tedras 873, 1563, 341, 323, Red HAWC, M\'{e}xico; DGAPA-UNAM grants IG101323, IN111716-3, IN111419, IA102019, IN106521, IN114924, IN110521 , IN102223; VIEP-BUAP; PIFI 2012, 2013, PROFOCIE 2014, 2015; the University of Wisconsin Alumni Research Foundation; the Institute of Geophysics, Planetary Physics, and Signatures at Los Alamos National Laboratory; Polish Science Centre grant, DEC-2017/27/B/ST9/02272; Coordinaci\'{o}n de la Investigaci\'{o}n Cient\'{i}fica de la Universidad Michoacana; Royal Society - Newton Advanced Fellowship 180385; Generalitat Valenciana, grant CIDEGENT/2018/034; The Program Management Unit for Human Resources \& Institutional Development, Research and Innovation, NXPO (grant number B16F630069); Coordinaci\'{o}n General Acad\'{e}mica e Innovaci\'{o}n (CGAI-UdeG), PRODEP-SEP UDG-CA-499; Institute of Cosmic Ray Research (ICRR), University of Tokyo. H.F. acknowledges support by NASA under award number 80GSFC21M0002. We also acknowledge the significant contributions over many years of Stefan Westerhoff, Gaurang Yodh and Arnulfo Zepeda Dom\'inguez, all deceased members of the HAWC collaboration. Thanks to Scott Delay, Luciano D\'{i}az and Eduardo Murrieta for technical support.
\end{acknowledgements}


\bibliography{bibliography}{}
\bibliographystyle{aasjournal}



\end{document}